# Recent advances in understanding and manipulating magnetic and electronic properties of Eu$M_2X_2$ ($M$ = Zn, Cd; $X$ = P, As)

Xiyu Chen,[1] Shuai Dong,[1] and Zhi-Cheng Wang[1,†]

[1]*Key Laboratory of Quantum Materials and Devices of Ministry of Education, School of Physics, Southeast University, Nanjing 211189, China*

**ABSTACT**

Over the past five years, significant progress has been made in understanding the magnetism and electronic properties of CaAl$_2$Si$_2$-type Eu$M_2X_2$ ($M$ = Zn, Cd; $X$ = P, As) compounds. Prior theoretical work and experimental studies suggested that EuCd$_2$As$_2$ had the potential to host rich topological phases, particularly an ideal magnetic Weyl semimetal state when the spins are polarized along the *c* axis. However, this perspective is challenged by recent experiments utilizing samples featuring ultra-low carrier densities, as well as meticulous calculations employing various approaches. Nonetheless, the Eu$M_2X_2$ family still exhibit numerous novel properties that remain to be satisfactorily explained, such as the giant nonlinear anomalous Hall effect and the colossal magnetoresistance effect. Moreover, Eu$M_2X_2$ compounds can be transformed from semiconducting antiferromagnets to metallic ferromagnets by introducing a small number of carriers or applying external pressure, and a further increase in the ferromagnetic transition temperature can be achieved by reducing the unit cell volume. These features make the Eu$M_2X_2$ family a fertile platform for studying the interplay between magnetism and charge transport, and an excellent candidate for applications in spintronics. This paper presents a comprehensive review of the magnetic and transport behaviors of Eu$M_2X_2$ compounds with varying carrier densities, as well as the current insights into these characteristics. An outlook for future research opportunities is also provided.

---

[†] wzc@seu.edu.cn

# 1. Introduction

The trigonal CaAl$_2$Si$_2$-type Eu$M_2X_2$ ($M$ = Zn, Cd; $X$ = P, As) were first synthesized decades ago [1,2]. Reports on their crystal structures and basic magnetic and transport properties revealed them to be Zintl phases with a narrow bandgap and an antiferromagnetic (AFM) ground state [3,4]. However, these Eu-based materials did not garner significant attention, save for the potential of antimonides in thermoelectric applications [5,6]. This situation remained until the recent surge in research on magnetic topological materials. Among the Eu$M_2X_2$ compounds, EuCd$_2$As$_2$ first attracted substantial interest due to its structural similarity to the Dirac semimetal (DSM) Cd$_3$As$_2$ and complex interplay of magnetism and band topology [7,8]. It exhibits an AFM transition at a Néel temperature ($T_N$) of 9.5 K, characterized by an A-type AFM structure [3,8,9]. Weyl semimetal state (WSM) or DSM state below $T_N$ was claimed, based on the different angle-resolved photoemission spectroscopy (ARPES) results [10-12]. Furthermore, theoretical and experimental investigations suggest that EuCd$_2$As$_2$ is an ideal candidate for a magnetic WSM, featuring a $c$-axis polarized state [11,13,14]. And a Weyl state may potentially be induced by ferromagnetic (FM) spin fluctuations in the paramagnetic (PM) phase [10]. Due to its strong spin-orbit coupling (SOC), the band structure of EuCd$_2$As$_2$ is highly sensitive to spin canting [15]. This sensitivity leads to the observation of a pronounced nonlinear anomalous Hall effect (NLAHE), attributed to the spin-dependent band structure and its associated momentum-space Berry curvature [8,16,17]. Given the plethora of exotic phenomena it exhibits, EuCd$_2$As$_2$ stands out as the most intensively studied member of the Eu$M_2X_2$ family [18-24]. However, the views about the nontrivial band topology of EuCd$_2$As$_2$ have been challenged by the recent experiments and theoretical calculations [25-28]. Earlier studies indicated that the temperature-dependent resistivity was predominantly metallic [8,9]. Yet, recent investigations utilizing single crystals with reduced carrier densities ($n_h \approx 10^{15}$ cm$^{-3}$) suggest that EuCd$_2$As$_2$ is indeed a magnetic semiconductor rather than a WSM [25-27].

Inconsistent properties are also observed across other members of the Eu$M_2X_2$ family. Wang $et$ $al$ reported that EuZn$_2$As$_2$ is an antiferromagnet with $T_N$ of 19.6 K, displaying a bad-metal behavior with a carrier density of $8.6 \times 10^{17}$ cm$^{-3}$ at 200 K, and shows a negative magnetoresistance (nMR) of −300% [magnetoresistance defined as MR = 100% × ($\rho(H) - \rho(0)$) / $\rho(H)$] near $T_N$ [29]. Crystals of EuZn$_2$As$_2$ grown by Blawat $et$ $al$ showed a significantly higher carrier density, approximately $4 \times 10^{20}$ cm$^{-3}$, yet exhibited a comparable nMR effect [30]. More recently, Luo $et$ $al$ succeeded in growing single crystals of EuZn$_2$As$_2$ with a notably lower carrier density ($n_h$ = $1.45 \times 10^{17}$ cm$^{-3}$ at 200 K) revealing a semiconducting nature below 100 K, accompanied by a colossal magnetoresistance (CMR) effect. And a topological phase transition is believed to be induced by pressure [31]. Intriguingly, despite the variations in transport behavior, the reported $T_N$ values for EuZn$_2$As$_2$ across these studies are nearly identical. The disparate transport properties can likely be attributed to differences in hole doping levels, associated with cation (Eu$^{2+}$) defects within the crystal lattice.

Regarding EuZn$_2$P$_2$ ($T_N$ = 23.5 K), notable discrepancies in reported transport behaviors have been observed. Berry $et$ $al$ described EuZn$_2$P$_2$ as exhibiting insulating behavior, with resistivity reaching magnitudes far exceeding $10^4$ Ω cm at 100 K [32]. Conversely, Krebber $et$ $al$ reported a resistivity of merely 0.06 Ω cm at the same temperature, alongside the observation of a CMR effect under magnetic field [33]. Moreover, EuCd$_2$P$_2$ ($T_N$ = 11 K), the Cd analogue of EuZn$_2$P$_2$, stands out as a distinct member within the Eu$M_2X_2$ family. Unlike its sibling compounds, EuCd$_2$P$_2$ shows a resistivity peak above $T_N$, attributed to pronounced magnetic fluctuations, whereas the peaks for other Eu$M_2X_2$ compounds are situated precisely at $T_N$. Wang $et$ $al$ reported a resistivity peak at 18 K and a remarkably high CMR effect, exceeding $10^4$%, upon suppression of the peak with a magnetic field [34]. By contrast, Zhang $et$ $al$ reported both the resistivity peak (at 14 K) and the CMR effect for EuCd$_2$P$_2$ to be several orders of magnitude larger [35].

Fundamentally, the diverse properties manifested by the same Eu$M_2X_2$ compound can be ascribed to the intrinsic sensitivity of its narrow electronic bandgap to variations in carrier concentration, coupled with the spin configurations at the Eu sites. In the cases discussed, the synthesis protocols significantly impact the carrier density within the crystals, thereby influencing the ultimate properties of Eu$M_2X_2$ materials. This variability presents opportunities to manipulate the properties of Eu$M_2X_2$ compounds through adjusting the doping level by changing crystal growth conditions, employing chemical doping strategies, or utilizing other techniques such as hydrostatic pressure and electrostatic gating. In fact, recent studies on Eu$M_2X_2$ ($M$ = Zn, Cd; $X$ = P, As) crystals grown using the molten salt flux method demonstrate that the application of this technique not only alters the electrical transport properties significantly, but also enables tuning of the magnetic orders from AFM to FM states [36-39]. This transformation is facilitated by the increased Eu defects introduced by the salt flux method, which is supported by the corresponding single crystal refinement data. The enhanced carrier concentration resulting from these defects induces interlayer FM interactions, leading to a FM ground state. Notably, this transition can be achieved with just a few percent of Eu vacancies or even less [38,39].

**Table 1.** The cell parameters, Néel temperatures ($T_N$), and Weiss temperatures ($\Theta_w$) of AFM-Eu$M_2X_2$ ($M$ = Zn, Cd; $X$ = P, As).

| Compound | $a = b$ (Å) | $c$ (Å) | $c/a$ | $V_{cell}$ (Å$^3$) | Eu $T_N$ (K) | $\Theta_w$ | Ref. |
|---|---|---|---|---|---|---|---|
| EuZn$_2$P$_2$ | 4.08497(18) | 7.0019(4) | 1.714 | 101.187(11) | 23.5 | 19.2 | [32] |
| EuZn$_2$As$_2$ | 4.21118(3) | 7.18114(6) | 1.705 | 110.2888(24) | 19.6 | 20.2 | [29] |
| EuCd$_2$P$_2$ | 4.3248(2) | 7.1771(7) | 1.660 | 116.26 | 11.3 | 28.1 | [34] |
| EuCd$_2$As$_2$ | 4.44016(4) | 7.32779(9) | 1.650 | 125.1125(38) | 9.2 | 12.1 | [29] |

Recent reports on the diverse physical properties underscore that the Eu$M_2X_2$ family ($M$ = Zn, Cd; $X$ = P, As) exhibits highly tunable magnetism and charge transport, positioning it as a promising material system for future technological applications. Therefore, it is imperative to review the advances in comprehending and manipulating the magnetic and transport properties of Eu$M_2X_2$ compounds. Firstly, we will succinctly outline the crystal structure, synthesis procedures, and the contradictory theoretical predictions surrounding Eu$M_2X_2$. Following that, we will contrast the physical properties of heavily hole-doped versus low-carrier-density AFM-Eu$M_2X_2$ crystals, encompassing magnetism, transport properties, pressure effects, and chemical doping. Subsequently, we will delve into the novel properties exhibited by FM-Eu$M_2X_2$ samples obtained through the salt flux method. Finally, we will summarize a phase diagram delineating the relationship between resistivity magnitude and carrier density, which clearly elucidates the interplay between magnetism and charge transport. Given the vast body of research on the Eu$M_2X_2$ family, attempting to provide comprehensive coverage is neither practical nor necessary; hence, this review selectively focuses on salient studies related to the property manipulation of Eu$M_2X_2$ materials. For instance, the Eu-based antimonides will not be specifically highlighted, since of the relative scarcity of research on property manipulation [40-42]. The overarching goal of this review is to present an overview of the varied properties of Eu$M_2X_2$ compounds, identify the primary factors influencing their magnetism and transport behaviors, and outline the experimental strategies required to effectively manipulate the properties of the Eu$M_2X_2$ family for advanced applications.

## 2. Crystal structure

Eu$M_2X_2$ ($M$ = Zn, Cd; $X$ = P, As) crystallizes in a CaAl$_2$Si$_2$-type structure (trigonal, space group $P$-3$m$1, No. 164), as depicted in figure 1. The Eu$^{2+}$ ions are arranged in a triangular lattice within the $ab$ plane. $M$ atoms occupy the centers of the $MX_4$ tetrahedra, and the tetrahedra form the quasi-two-dimensional [$M_2X_2$]$^{2-}$ frameworks through edge-sharing. The layers of Eu$^{2+}$ ions are connected by the anionic [$M_2X_2$]$^{2-}$ slabs. Indeed, the CaAl$_2$Si$_2$-type structure is prevalent among ternary Zintl compounds containing rare-earth or alkaline-earth elements [43]. Concerning Eu-based compounds, over a dozen phases have been identified, exemplified by EuAl$_2$Ge$_2$, EuMg$_2$Bi$_2$ and EuMn$_2$As$_2$, with many exhibiting an A-type AFM ordering [44-46]. This review specifically addresses the Zn and Cd variants, whose crystallographic parameters listed in Table 1.

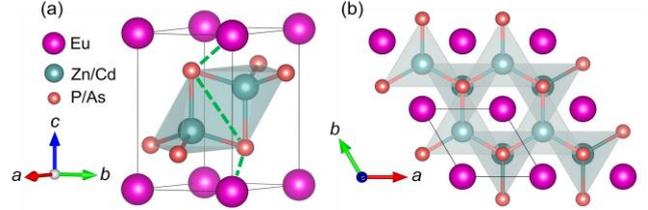

**Figure 1.** The crystal structure of Eu$M_2X_2$ ($M$ = Zn, Cd; $X$ = P, As). Two distinct views of the crystal structure are presented in panels (a) and (b), respectively. The green dashed line shows the path for the interlayer superexchange interaction.

## 3. Crystal growth methods

Single crystals of Eu$M_2X_2$ can be grown using the metal flux or molten salt flux, both of which are detailed in this section. These differing growth techniques result in crystals with variable carrier densities. In fact, before the extensive investigations on CaAl$_2$Si$_2$-type Eu$M_2X_2$ compounds, it was recognized that defects at the cation sites can significantly impact electronic properties, as evidenced by studies on $A$Zn$_2$Sb$_2$ ($A$ = Sr, Ca, Yb, Eu) as prospective thermoelectric materials [47-49]. The concentrations of $A$-site cation vacancies are strongly dependent on the cation electronegativity and sample growth conditions. The primary focus of this review will be on elucidating the vacancy-controlled magnetic and charge transport properties of Eu$M_2X_2$. It is noteworthy to emphasize that, although the diverse crystal growth techniques yield samples with distinct transport properties, the crystal structure remains essentially unchanged. The main alteration discerned across these crystals manifest in the variation of carrier concentration.

*Conventional metal flux method.* The metal flux method is a common way to grow the crystals of Eu$M_2X_2$, utilizing Sn (or Bi) as the flux [10,50]. The density of intrinsically formed cation vacancies is contingent upon the purity of the initial materials and the specific crystal synthesis process. Despite variations, the procedures outlined in various publications are largely the same: the elements are mixed with Sn at a ratio Eu$M_2X_2$: Sn = 10~30, then heated to a high temperature between 800~1150°C for an extended duration from several hours to 36 hours. Following this, the mixture undergoes a gradual cooldown at a rate of 2~3°C per hour before the crystals are separated via centrifugation. However, the resultant hole concentrations and properties of Eu$M_2X_2$ crystals, can exhibit considerable variability when produced through this method. As an illustrative case, EuCd$_2$As$_2$ has been reported to yield both insulating samples with $n \sim 10^{13}$ cm$^{-3}$ and metallic samples with $n \sim 10^{19}$ cm$^{-3}$ [10,27]. Despite the diversity in charge transport characteristics observed across these samples, their ground states intriguingly converge on antiferromagnetism, marked by a consistent $T_N$.

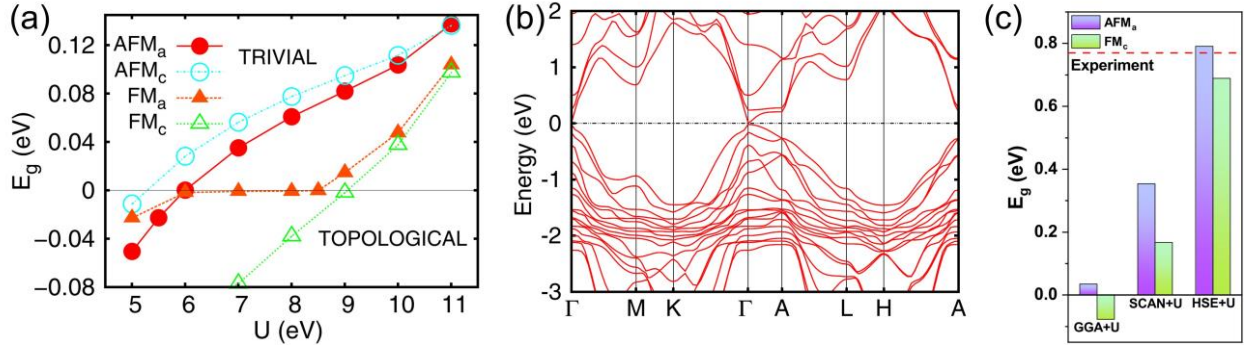

**Figure 2.** (a) Energy gap ($E_g$) for EuCd$_2$As$_2$ as a function of the Coulomb repulsion energy $U$. AFM$_a$: AFM phase with spins in the $ab$ plane; AFM$_c$: AFM phase with spins along the $c$ axis; FM$_a$: FM phase with spins in the $ab$ plane; FM$_c$: FM phase with spins along the $c$ axis. The negative values of $E_g$ indicate a topological phase, whereas positive values represent a trivial phase. (b) Band structure EuCd$_2$As$_2$ in the A-type AFM configuration of by using GGA + $U$ with $U$ = 7eV on the $f$ orbitals of Eu. (c) $E_g$ obtained within GGA + $U$, SCAN + $U$, HSE + $U$ ($U$ = 7 eV) for the AFM$_a$ and FM$_c$ phases. Reprinted from [28].

*Modified Sn flux method.* We notice that the recipes for the Sn flux method, as detailed in several publications, deviate subtly from the conventional procedure. Santos-Cottin *et al* adopted a two-step process to grow EuCd$_2$As$_2$ crystals, where the initial growth products served as seed material for the final growth, thereby increasing the crystal size and quality [25]. Through this approach, they obtained EuCd$_2$As$_2$ samples featuring a bandgap of about 770 meV. In another study about EuZn$_2$P$_2$, Krebber *et al* used a graphite crucible rather than the alumina crucible as the container for the starting materials, which yields samples with significantly lower resistivity compared to those reported elsewhere [33]. Although the specific impact of the graphite crucible was not explicitly elucidated, a plausible hypothesis is that the subtle replacement of phosphorus with carbon may underlie the enhanced electrical conductivity. Interestingly, Usachov *et al* synthesized crystals of EuCd$_2$P$_2$ using the graphite crucible [51], revealing a complex magnetic behavior and a more pronounced CMR effect compared to the sample in Ref. [34].

*Salt flux method.* In the early stages, Schellenberg *et al* synthesized EuCd$_2$As$_2$ single crystals using the salt flux method, with a molar ratio of EuCd$_2$As$_2$ to an equimolar mixture of NaCl/KCl set at 1:4, which yielded crystals displaying an AFM ground state [3]. Later, Jo *et al* achieved the first successful growth of EuCd$_2$As$_2$ crystals with a FM state using the same salt flux method, albeit with a mass ratio of EuCd$_2$As$_2$ to the salt mixture adjusted to 1:4, translating to a molar ratio of 1:16 [36]. This suggests that the FM sample was grown in a more dilute solution. Furthermore, Jo *et al* also synthesized AFM-EuCd$_2$As$_2$ crystals in the dilute salt flux environment by increasing the proportion of Eu, setting the ratio to Eu:Cd:As = 1.75:2:2 [36]. Both the use of a concentrated flux and the heightened proportion of Eu indicate that AFM-EuCd$_2$As$_2$ crystals should possess fewer Eu vacancies than their FM counterparts, which is supported by crystal structural refinements. Very recently, our team successfully synthesized EuCd$_2$P$_2$, EuZn$_2$As$_2$, and EuZn$_2$P$_2$ crystals with FM ground states through the salt flux method [38,39]. This achievement underscores the versatility of the salt flux method as a general strategy for modulating the ground state of CaAl$_2$Si$_2$-type Eu$M_2X_2$ compounds.

## 4. Electronic structure

EuCd$_2$As$_2$ was initially deemed a promising candidate for hosting rich topological phases, which was supported by a plethora of experimental and theoretical investigations [8,10-14]. However, recent studies on high-quality EuCd$_2$As$_2$ crystals, characterized by reduced carrier densities, have revealed EuCd$_2$As$_2$ is actually a magnetic semiconductor rather than the anticipated topological WSM [25-27]. Moreover, recent calculations by Cuono *et al* have underscored the significant influence of electron correlation effects on the electronic and topological properties of EuCd$_2$As$_2$ [28]. Previous theoretical studies, employing modest Hubbard $U$ values (~ 3-5 eV) within the standard generalized gradient approximation (GGA) functional, typically concluded that EuCd$_2$As$_2$ exhibits nontrivial topology. In contrast, computations utilizing the Heyd-Scuseria-Ernzerhof (HSE) hybrid functional + $U$ and the strongly constrained and appropriately normed (SCAN) functional + $U$ approaches tend to yield positive values of the energy gap, indicating a trivial band topology for EuCd$_2$As$_2$, which aligns more closely with experimental findings [25]. Shi et al further confirmed the underestimation of the Hubbard $U$ in earlier studies by employing the linear response ansatz, determining a converged $U$ value of approximately 8 eV [26]. In this section, we will first briefly summarize the statements from prior research that claimed the topologically nontrivial nature of EuCd$_2$As$_2$, followed by an exposition of the revised conclusions from Cuono et al [28] and associated insights into the electronic structure of Eu$M_2X_2$ compounds.

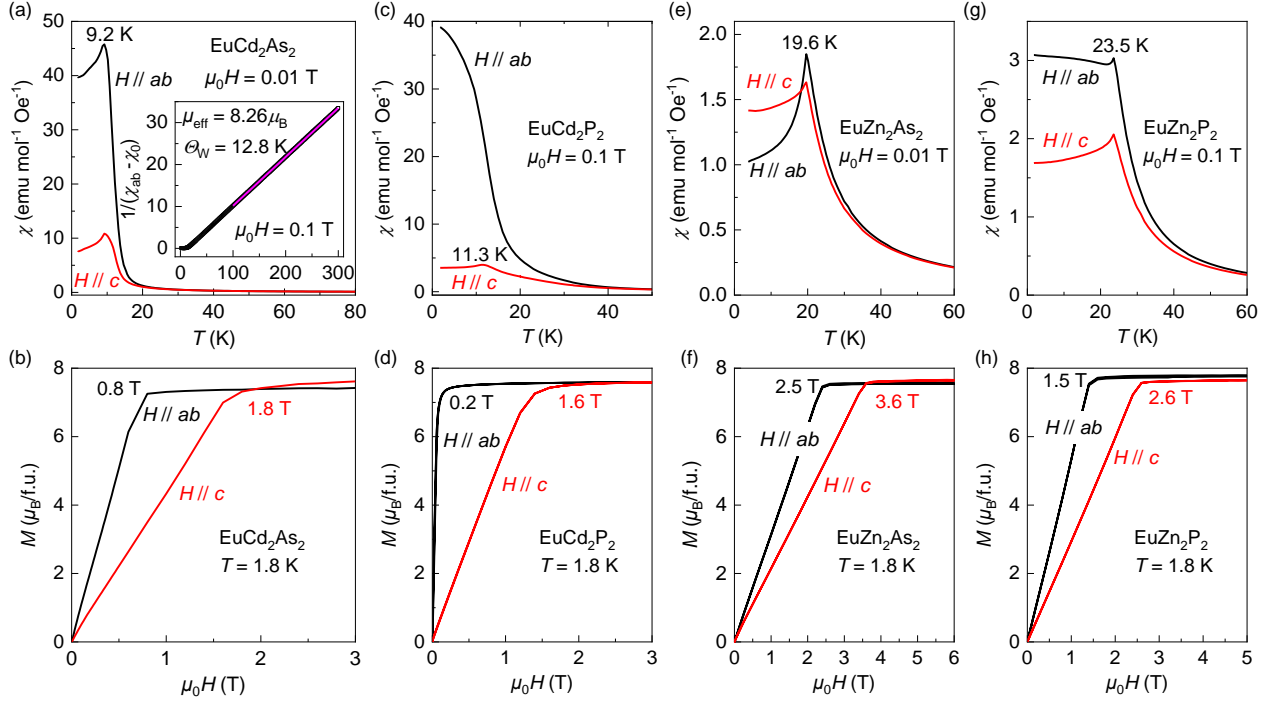

**Figure 3.** Temperature-dependent magnetic susceptibility of antiferromagnetically ordered EuCd$_2$As$_2$ (a), EuCd$_2$P$_2$ (c), EuZn$_2$As$_2$ (e), and EuZn$_2$P$_2$ (g) under both in-plane and out-of-plane fields. The inset of panel (a) displays the Curie-Weiss fit of EuCd$_2$As$_2$. Correspondingly, the in-plane and out-of-plane magnetization data at 1.8 K for each compound are illustrated in panels (b), (d), (f), and (h), respectively. Panels [(a), (b), (e), (f)] are reproduced from [29]. Panels [(c), (d)] are reproduced from [34].

Hua et al predicted EuCd$_2$As$_2$ would host a DSM state when the ground state assumes an A-type AFM structure with Eu spin aligned along the $c$ axis (AFM$_c$) [14]. However, when the AFM structure features an in-plane spin configuration (AFM$_a$), the threefold rotation symmetry $C_3$ is broken, transforming EuCd$_2$As$_2$ to an AFM topological insulator. In a concurrent experimental study, Rahn et al revealed that the actual magnetic ground state of EuCd$_2$As$_2$ is AFM with spins lying in the $ab$ plane [8]. Subsequently, Soh et al predicted the existence of a WSM phase, characterized by a single pair of Weyl nodes near the Fermi level, under the condition that Eu spins are fully polarized along the $c$ axis [11]. This prediction was supported by their quantum oscillation and ARPES results. Wang et al reported a similar conclusion and further predicted that substituting half of the Eu sites with Ba could stabilize FM order [13]. The experimental results obtained by Ma et al showed quasi-static and quasi-long-range FM fluctuations in the PM phase of EuCd$_2$As$_2$, which could lead to band splitting and induce a WSM state [10]. The following year, Ma et al also claimed that a magnetic DSM-like band structure was observed via ARPES in the AFM state of EuCd$_2$As$_2$ [12].

These investigations appear robust enough to substantiate the nontrivial band topology of EuCd$_2$As$_2$. However, an optical spectroscopy study conducted by Santos-Cottin et al, utilizing low-carrier-density EuCd$_2$As$_2$, indicates that it is in fact a magnetic semiconductor with an optical gap of 0.77 eV [25]. Furthermore, the semiconducting EuCd$_2$As$_2$ sample grown by Shi et al exhibits a similar optical gap of 0.74 eV and a transport gap of 55.8 meV [26]. Electrical transport results of semiconducting EuCd$_2$As$_2$ will be presented in Section 5.2. Directly following Santos-Cottin et al's work, Cuono et al calculated the electronic properties of EuCd$_2X_2$ ($X$ = P, As, Sb, Bi) using the HSE + $U$, SCAN + $U$, and GGA + $U$ approaches [28]. Within GGA + $U$, they found that the sign of the band gap is highly sensitive to variations in the value of $U$, as illustrated in figure 2(a) [28]. For the AFM$_a$ ground state, a transition occurs from the topological to the trivial phase at $U$ = 6.0 eV. When $U$ = 7 eV with AFM$_a$ spin configuration, EuCd$_2$As$_2$ is in a trivial phase, featuring a notably narrow band gap of 40 meV [28]. The corresponding band structure is displayed in panel (b). Cuono et al compared the resulting $E_g$ value from the GGA + $U$ to those derived from HSE + $U$ and SCAN + $U$ in panel (c). They found that the latter two approaches yield much larger $E_g$ values, and HSE + $U$ provides a gap value of 0.79 eV that is highly consistent with experimental data [25,26,28]. Generally, GGA + $U$ approach tends to underestimate the band gap due to its limitations, while computations utilizing hybrid functionals produce more accurate band gaps. Therefore, the excellent consistency of the gap values obtained from optical measurements and computation within HSE + $U$ clearly indicates the semiconducting nature of EuCd$_2$As$_2$. An additional key insight from their computations is that the FM$_c$ phase exhibits a smaller gap than the AFM phases across all $U$ values. Consequently, applying a magnetic field along the $c$ axis invariably diminishes the band gap, which resonates with experimental observations in Eu$M_2X_2$ materials and partially elucidates their nMR effect.

The trivial band gap of EuCd$_2$P$_2$, as computed by Cuono *et al*, is approximately 1.4 eV [28]. While the $E_g$ values reported by Zhang *et al* and Chen *et al* are 819 meV and 380 meV, respectively, due to their use of different functionals and smaller Hubbard $U$ values [35,39]. Chen *et al* also compared the total energies of EuCd$_2$P$_2$ with and without Eu vacancies, under both AFM and FM spin configurations. Their analysis revealed that the interlayer FM coupling is more favored in the presence of Eu vacancies, which is consistent with the experimental observations [39]. Moreover, both EuZn$_2$As$_2$ and EuZn$_2$P$_2$, when in the AFM$_a$ state, are calculated to be trivial narrow-gap magnetic semiconductors [29,31-33,52].

## 5. Physical properties of AFM-Eu$M_2X_2$

### 5.1 Magnetic properties

As previously mentioned, the magnetism in Eu$M_2X_2$ compounds originates from the triangular lattice of Eu$^{2+}$ ions with a 4$f^7$ electronic configuration. All members of Eu$M_2X_2$ ($M$ = Zn, Cd; $X$ = P, As) exhibit an A-type AFM order, implying that the Eu$^{2+}$ spins couple ferromagnetically within the $ab$ plane, whereas the interlayer interaction between Eu layers is AFM. Magnetic data for Eu$M_2X_2$ are summarized in figure 3. Considering the similarity of the magnetic behavior, we use the data from EuCd$_2$As$_2$, shown in panels (a) and (b), as a representative example [9,29]. Panel (a) reveals a transition peak at 9.2 K for both in-plane and out-of-plane magnetic susceptibility curves, indicating the emergence of AFM order in EuCd$_2$As$_2$. A notable disparity in susceptibility with the magnetic fields in different directions underscores the pronounced magnetocrystalline anisotropy in EuCd$_2$As$_2$, with the $ab$ plane serving as the easy magnetization plane. A positive Weiss temperature ($\Theta_w$ = 12.1 K) is derived from fitting the Curie-Weiss law $\chi_{ab} = \chi_0 + C/(T - \Theta_w)$ to the susceptibility data above 100 K, suggesting dominant FM interactions. Panel (b) shows the magnetization of EuCd$_2$As$_2$ for different field directions. Saturation fields for in-plane and out-of-plane directions are determined to be 0.8 T and 1.6 T, respectively, confirming the easy-plane anisotropy. The saturated moment (7.5 $\mu_B$) under high fields and the effective moment of (8.3 $\mu_B$) obtained from the Curie-Weiss fit corroborate the Eu$^{2+}$ oxidation state.

EuCd$_2$P$_2$, EuZn$_2$As$_2$, and EuZn$_2$P$_2$ exhibit similar magnetic characteristics akin to those of EuCd$_2$As$_2$, with respective A-type AFM ordering temperatures of 11.3 K, 19.6 K, and 23.5 K, as shown in panels (c), (e), and (g) [29,34]. The shared features across these compounds include: (1) The A-type AFM order, substantiated through neutron diffraction or resonant elastic x-ray scattering (REXS) [8,29,33,53,54]; (2) The same easy magnetization plane (the $ab$ plane); (3) The positive Weiss temperatures, as listed in Table 1. Despite these commonalities, the differences in magnetocrystalline anisotropy are noticeable. EuCd$_2$P$_2$ shows a more pronounced anisotropy relative to EuCd$_2$As$_2$, evidenced by greater disparities in anisotropic susceptibilities and the conspicuous absence of an AFM peak in the in-plane susceptibility curve. This enhanced magnetic anisotropy stems from the reduced spatial extension of the $p$ orbitals from As to P, leading to weakened interlayer coupling between Eu layers [34]. The weakened interlayer coupling further results in the intensified magnetic fluctuations and a resistivity peak above $T_N$, as shown in figure 6(a). In contrast, EuZn$_2$As$_2$ and EuZn$_2$P$_2$ display milder magnetic anisotropy, attributable to the more localized $d$ orbitals and diminished SOC in the Zn compounds [29,32,52]. In the Eu$M_2X_2$ family, FM fluctuations above $T_N$ are commonly observed, playing a crucial role in facilitating certain exotic phenomena, such as the NLAHE [16,55].

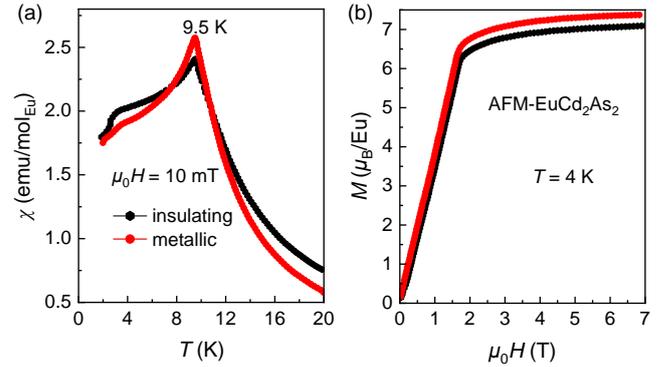

**Figure 4.** Comparisons of the $\chi(T)$ and $M(H)$ curves between metallic and insulating EuCd$_2$As$_2$. Reproduced from [25].

The ordering temperature of the Eu lattice depends on the specific $M_2X_2$ framework. Berry *et al* claimed that the magnetic interactions originate from the dipolar interactions, and that $T_N$ values can be scaled through a linear combination of $1/d_{nn}^3$ ($d_{nn}$ = intralayer Eu-Eu distance) and $1/d_{il}^3$ ($d_{il}$ = interlayer Eu-Eu distance), with the coefficients close to a 1:−4 ratio [32]. However, Singh *et al* explained the magnetic interactions with an extended superexchange mechanism, which was supported by their simulation results using Monte Carlo method [52]. Based on the scheme proposed by Singh *et al*, the interlayer AFM coupling between two Eu sites along the $c$ axis is mediated by two intervening pnictogen atoms, as illustrated by the green dashed line in figure 1(a). This superexchange mechanism effectively elucidates the variation in $T_N$ across Eu$M_2X_2$ compounds, taking into account the length of the Eu-$X$-$X$-Eu path. Among the quartet, EuCd$_2$As$_2$, featuring the longest Eu-As-As-Eu path (~10.81 Å), consequently exhibits the lowest $T_N$. Conversely, EuZn$_2$P$_2$, with the shortest path (~10.12 Å), presents the highest $T_N$.

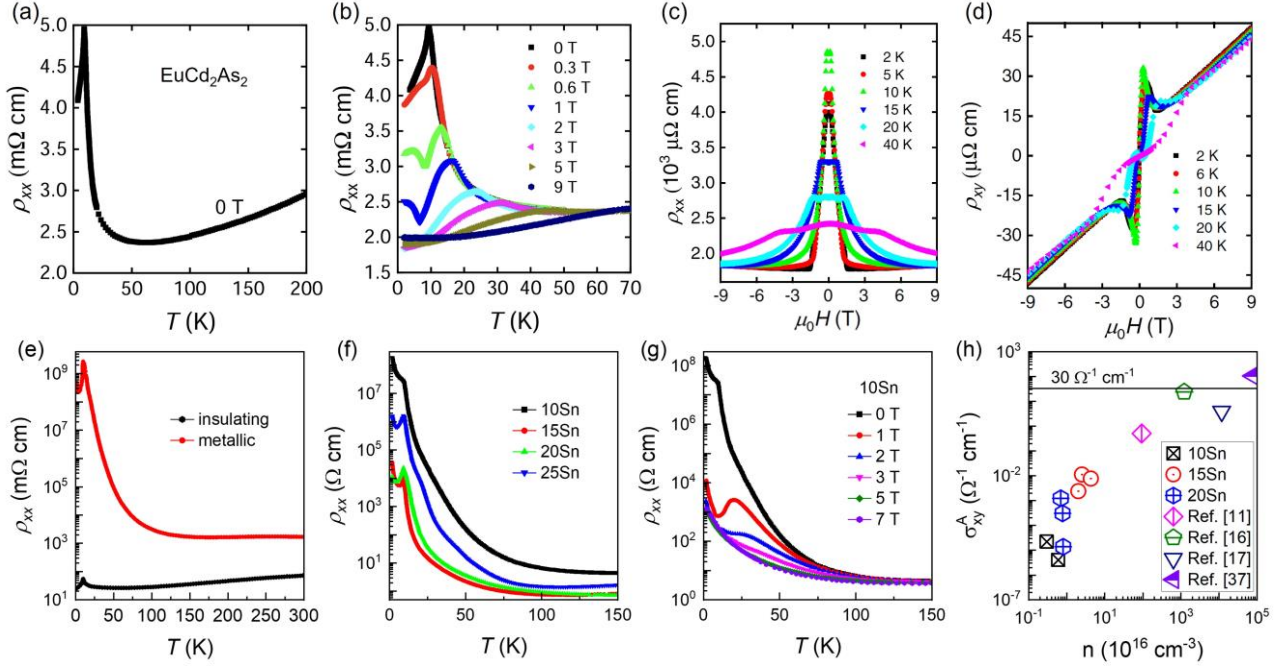

**Figure 5.** Electrical transport results of AFM-EuCd$_2$As$_2$ with metallic (a-d) (reprinted from [17]) and insulating (f-h) behaviors (reproduced from [26]). (a) Temperature-dependent resistivity, $\rho_{xx}(T)$, of metallic EuCd$_2$As$_2$ under zero field. (b) $\rho_{xx}(T)$ under different fields. (c) The field-dependent $\rho_{xx}$ measured at various temperatures. (d) Hall resistivity $\rho_{xy}$ of metallic EuCd$_2$As$_2$. (e) Comparison between $\rho_{xx}(T)$ of metallic and insulating EuCd$_2$As$_2$. Reprinted from [25]. (f) $\rho_{xx}(T)$ of samples grown with the different ratios of Sn flux. (g) $\rho_{xx}(T)$ under different fields from the sample grown with 10Sn. (h) Anomalous Hall conductivity $\sigma_{xy}^A$ versus carrier density $n$ of samples from different sources [11,16,17,26,37].

Moreover, it appears that the $T_N$ values and other magnetic properties of Eu$M_2X_2$ do not fluctuate significantly with changes in carrier density of the crystals, provided that the ground state is AFM. This observation is substantiated by the consistent results from various independent studies. As shown in figure 4, comparisons of susceptibility and magnetization between the insulating and metallic phases of EuCd$_2$As$_2$ reveal negligible discrepancies [25]. The underlying rationale for this phenomenon could be that the AFM interactions, specifically the superexchange correlations in Eu$M_2X_2$, predominantly hinge on the crystal structure. And a minor fraction of vacancies does not substantially modify the structure. However, when the FM interactions, triggered by the heavily doped carriers, become predominant, the ground state of Eu$M_2X_2$ undergoes a sudden switch, which will be discussed further in Section 6.2.2.

### 5.2 Electrical transport properties

In contrast to the relatively uniform magnetic properties, the electrical transport behaviors of compounds within the Eu$M_2X_2$ family exhibit a far greater diversity and are acutely sensitive to variations in carrier concentration. In this section, we delve into comparisons between the same AFM-Eu$M_2X_2$ materials with differing carrier concentrations. And we will sequentially introduce the charge transport properties of AFM-EuCd$_2$As$_2$, AFM-EuCd$_2$P$_2$, AFM-EuZn$_2$As$_2$, and AFM-EuZn$_2$P$_2$.

#### 5.2.1 Resistivity of AFM-EuCd$_2$As$_2$.
Figures 5(a-d) shows the electrical transport properties of EuCd$_2$As$_2$ reported by Xu *et al* [17]. The longitudinal resistivity, $\rho_{xx}$, decreases as temperature goes down, and this metallic behavior persists above 50 K. Below 50 K, enhanced magnetic scattering triggers an increase in $\rho_{xx}$, peaking at $T_N$ around 9.5 K, before declining again due to reduced scattering from the ordered Eu moments, as shown in figure 3(a). The resistivity peak at 9.5 K is notably diminished under the application of external magnetic fields. Figure 5(c) illustrates the $\rho_{xx}$ versus $H$

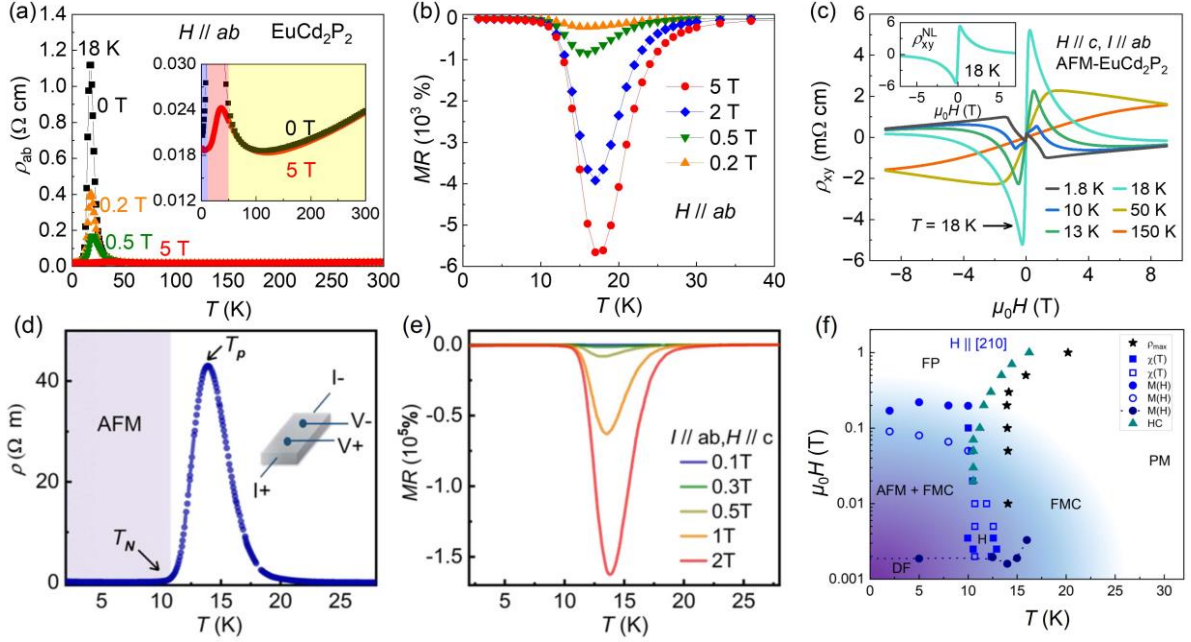

**Figure 6.** (a) Temperature-dependent resistivity of AFM-EuCd$_2$P$_2$ under various fields. Inset zooms in on the resistivity under 5 T. Reproduced from [34]. (b) Temperature dependence of MR, from [34]. (c) Hall resistivity of AFM-EuCd$_2$P$_2$ as a function of field at different temperatures. Inset shows the NLAHE at 18 K. Reproduced from [39]. [(d), (e)] Temperature-dependent resistivity and MR of AFM-EuCd$_2$P$_2$ with a lower carrier density. Reprinted from [35]. (f) $H$-$T$ phase diagram of AFM-EuCd$_2$P$_2$ for H // [210]. DF: domain flip; FP: field-polarized; PM: paramagnetic; FMC: FM clusters. Reprinted from [51].

dependence at various temperatures. At 2 K, a nMR of approximately −300% is attained at $\mu_0 H \sim 1.5$ T with the definition that MR = 100% × [$\rho_{xx}(H) - \rho_{xx}(0)$]/$\rho_{xx}(H)$. $\rho_{xx}$ reaches a minimum at 1.5 T, aligning with the saturation field of out-of-plane magnetization. Above $T_N$, a slight increase in $\rho_{xx}(H)$ is observed in the low-field region, which may be attributed to the field-induced canting of the Eu spin towards the $c$ axis.

The Hall resistivity, $\rho_{xy}$, of EuCd$_2$As$_2$ is shown in figure 5(d). The total Hall resistivity can be decomposed as $\rho_{xy} = R_0\mu_0H + S_H\rho_{xx}^2M + \rho_{xy}^{NL}$. The first term, $R_0\mu_0H$, represents the ordinary Hall effect (OHE), where $R_0$ denotes a constant. The conventional anomalous Hall effect (AHE) is represented by the second term, $S_H\rho_{xx}^2M$, with $S_H$ being another constant. The final component, $\rho_{xy}^{NL}$, signifies the unconventional contribution to the AHE, i.e., NLAHE. As shown in figure 5(d), it is evident that $\rho_{xy}$ varies markedly above and below $T_N$, which is influenced by differing contributions from $\rho_{xy}^{NL}$. Xu et al attributed this variation to the coexistence of two mechanisms below $T_N$: the real-space Berry phase driven topological Hall effect (THE) and the momentum-space Berry curvature associated with Weyl points. In contrast, only the latter mechanism contributes to $\rho_{xy}^{NL}$ above $T_N$ [17]. Cao et al also reported a similar giant NLAHE in EuCd$_2$As$_2$, and explored its physical origin [16]. However, they claimed that the NLAHE in EuCd$_2$As$_2$ does not stem from the real-space THE or Weyl physics, but rather arises from the evolution of the band structure induced by spin rotation [16]. This explanation is grounded in the extreme sensitivity of the electronic structure to spin canting.

Recently, Santos-Cottin *et al* obtained ultraclean EuCd$_2$As$_2$ crystals through a two-step growth process [25]. These crystals exhibit insulating transport characteristics, yet their magnetic properties are virtually indistinguishable from those of metallic EuCd$_2$As$_2$ samples, as illustrated in figure 4. Figure 5(e) compares the resistivity of the insulating phase to that of the metallic phase, revealing discrepancies exceeding seven orders of magnitude at low temperatures. Based on experimental evidence from electronic transport, optical spectroscopy, and excited-state photoemission spectroscopy, Santos-Cottin *et al* proposed that the ultraclean EuCd$_2$As$_2$ is actually a magnetic semiconductor with a gap of about 770 meV, rather than a field-induced topological semimetal. Their results also indicate that the band gap of EuCd$_2$As$_2$ can be notably reduced in the magnetic field [25].

Subsequently, several investigations have further confirmed the semiconductor nature of EuCd$_2$As$_2$. Shi *et al* synthesized a series of semiconducting EuCd$_2$As$_2$ samples by adjusting the ratio of Sn flux, yielding carrier concentrations ranging from $10^{15}$ to $10^{16}$ cm$^{-3}$ in the crystals [26]. The resistivity curves from these samples are presented in figures 5(f) and 5(g). Similar to the metallic phase, the CMR effect is also observed in the semiconducting phase, with a significantly higher magnitude, as shown in figure 5(g).

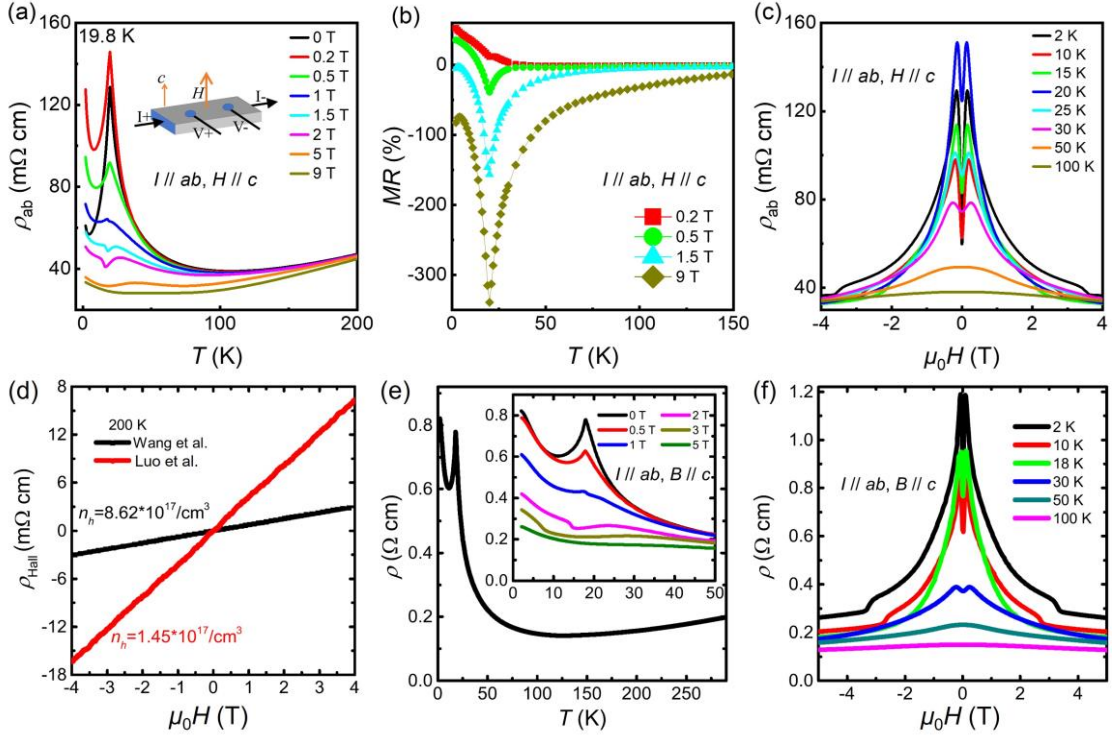

**Figure 7.** Electrical transport data of AFM-EuZn$_2$As$_2$ crystals with different carrier concentrations. Panels (a-c) shows the data from the sample with the higher carrier density, reprinted from [29], while panels (d-f) are reprinted from [31] to show the results of the sample with the lower carrier density. [(a), (e)] Temperature dependence of in-plane resistivity with fields along the $c$ axis. (b) Temperature dependence of MR under different fields. [(c), (f)] Field dependence of the resistivity at several temperatures. (d) A comparison between the Hall effects from the two samples at 200 K.

Through calculating the anomalous Hall conductivities (AHC) in low-carrier-density EuCd$_2$As$_2$ and juxtaposing these against the metallic samples, Shi *et al* discovered a strong correlation between the AHC of EuCd$_2$As$_2$ and the carrier density, as illustrated in figure 5(h). This finding contrasts with the anticipated constant AHC derived from the Berry curvature associated with the separation of the Weyl nodes [56]. Moreover, Shi *et al* attributed the prior misidentification of the topological phase in EuCd$_2$As$_2$, as claimed in numerous theoretical studies, to an underestimation of the Hubbard $U$ parameter [26].

In addition, Wang *et al* also carried out magneto-transport measurements on low-carrier-density ($10^{13}$ cm$^{-3}$) EuCd$_2$As$_2$ crystals [27]. They found that the surface conduction dominates the transport due to band bending, which is diminished upon mechanical polishing of the crystal surfaces. Moreover, the resistance can be dramatically modulated by applying a *dc* bias current due to Fermi surface displacement in electric fields. This investigation further challenges the validity of the topological semimetal hypothesis about EuCd$_2$As$_2$.

*5.2.2 Resistivity of AFM-EuCd$_2$P$_2$.* Among the Eu$M_2X_2$ compounds, EuCd$_2$P$_2$ was the first member to come to attention for exhibiting the CMR effect. Wang *et al* and Chen *et al* measured the in-plane resistivity and Hall effect of AFM-EuCd$_2$P$_2$ crystals, respectively [34,39]. The samples are sourced from the same batch, and the results are summarized in figures 6(a-c), whose carrier density in these crystals is estimated to be ~ $3.6 \times 10^{18}$ cm$^{-3}$ at 200 K [39]. A pronounced resistivity peak is observed at 18 K under zero field, well above the $T_N$ (11.3 K) for AFM-EuCd$_2$P$_2$. The peak is dramatically suppressed by a small field, leading to a giant CMR, as illustrated in figure 6(a). Three distinct temperature ranges are identified and highlighted in the inset: a yellow-shaded poor metallic region, a red-shaded intermediate CMR region, and a blue-shaded magnetic ordered region below $T_N$. The CMR maximum is situated at the temperature of the resistivity peak ($T_{peak}$), specifically within the red-shaded area, setting it apart from other Eu$M_2X_2$ compounds where the temperature of the maximum nMR coincides with $T_N$. The calculated MR exceeds $-10^3$% in fields less than 1 T, as shown in figure 6(b) plotted against temperature. Wang *et al* ascribe the CMR of EuCd$_2$P$_2$ to strong FM fluctuations above $T_N$. Moreover, the Hall response of EuCd$_2$P$_2$ near the resistivity peak temperature is entirely dominated by the NLAHE by AFM-EuCd$_2$P$_2$, as demonstrated in figure 6(c). A similar decomposition analysis of the Hall resistivity, akin to that performed for EuCd$_2$As$_2$ (figure 5), was conducted. This analysis revealed that the contribution of the NLAHE can reach up to 97% at 0.25 T for the curve at 18 K [39]. Given the topologically trivial band structure of EuCd$_2$P$_2$, the NLAHE may originate from a nonzero spin chirality related to the FM domains above $T_N$ [54] or the evolution of the electronic structure induced by the external magnetic field [16]. Further experimental clarifications are needed to explain the NLAHE in EuCd$_2$P$_2$.

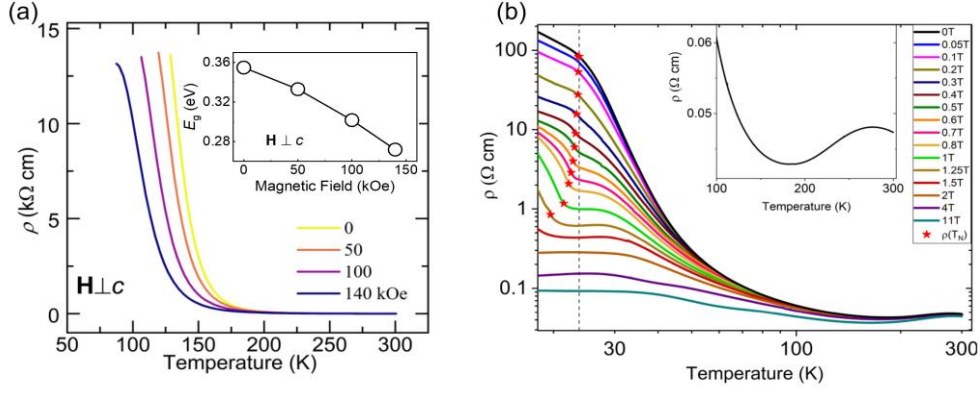

**Figure 8.** (a) In-plane resistivity of insulating AFM-EuZn$_2$P$_2$ in different fields transverse to the $c$ axis. Inset shows the band gap as a function of the magnetic field. Reproduced from [52]. (b) In-plane resistivity of semiconducting EuZn$_2$P$_2$ in different fields along the $c$ axis. Inset shows the resistivity in zero field above 100 K. Reprinted from [33].

The origin of CMR in EuCd$_2$P$_2$ has attracted considerable research interest, as it clearly cannot be attributed to mechanisms analogous to those in manganites, given the absence of mixed valence or significant lattice distortion. Flebus *et al* posited a magnetic Berezinskii-Kosterlitz-Thouless transition as the explanation for the pronounced CMR [57,58]. Furthermore, Homes *et al* and Sunko *et al* reported the emergence of FM clusters above the $T_N$, leading to carrier localization and an increase in resistivity via spin-carrier interactions [54,59]. Zhang *et al* carried out ARPES to investigate the origin of CMR [35]. Their findings suggested an electronic structure reconstruction, transitioning from PM (above $T_{peak}$) to FM (between $T_{peak}$ and $T_N$), and finally to AFM (below $T_N$). It is worth noting that the EuCd$_2$P$_2$ sample analyzed by Zhang *et al* exhibits characteristics distinct from those studied by Wang *et al*, as manifested by a resistivity peak and CMR effect approximately ten times greater, as illustrated in figures 6(d,e), along with a lower $T_{peak}$ (14 K), indicative of a reduced carrier concentration in the crystal [34,35]. In a recent study, Usachov *et al* examined the EuCd$_2$P$_2$ crystals synthesized using a graphite crucible [51]. Their electrical transport measurements also revealed a resistivity peak at 14 K rather than 18 K, yet the peak magnitude and CMR effect were comparable to the outcomes reported by Wang *et al* [34]. Contrary to Zhang *et al*'s observations [35], Usachov *et al*'s spectroscopic data showed no evidence of exchange splitting or substantial changes in the electronic structure [51]. Additionally, they constructed a $H$-$T$ phase diagram of EuCd$_2$P$_2$ using magnetization, heat capacity, and transport data, as depicted in figure 6(f). Further experimentation is warranted to reconcile the discrepancies between different reports and elucidate the mechanisms underlying CMR phenomena above $T_N$ in EuCd$_2$P$_2$.

*5.2.3 Resistivity of AFM-EuZn$_2$As$_2$.* Several teams have investigated the charge transport properties of AFM-EuZn$_2$As$_2$, yielding largely consistent results except for the carrier densities within the crystals and the absolute values of resistivity [29-31,55]. EuZn$_2$As$_2$ was predicted to be a narrow-gap semiconductor by first-principles calculations [29]. However, it displays a bad-metal behavior at high temperatures (> 100 K), as evidenced by the results from Wang *et al* (figures 7(a)-(c)) and Luo *et al* (figures 7(d)-(f)) [29,31]. Analogous to the case of AFM-EuCd$_2$As$_2$, a resistivity peak at $T_N$ and an increase in resistivity above $T_N$ due to short-range magnetic correlations have been observed. Magnetic fluctuations above $T_N$ are suppressed by applied external fields, leading to a nMR around $T_N$. Wang *et al*'s data reveal that this nMR peak reaches a maximum of −340 % at $T_N$ with an out-of-plane field of 9 T [29]. Figure 6(c) illustrates that Wang *et al*'s sample ($n \sim 8.62 \times 10^{17}$ cm$^{-3}$) possesses a higher hole concentration compared to Luo *et al*'s ($n \sim 1.42 \times 10^{17}$ cm$^{-3}$). Despite this difference, the magnitudes of nMR (figures 7(a) and 7(e)) and the field-dependence of resistivity (figures 7(c) and 7(f)) are quite similar. Yi *et al* also reported a comparable behavior in EuZn$_2$As$_2$, albeit with resistivities approximately 700 times greater [55]. Moreover, Blawat *et al* discovered a pronounced angular dependence of MR at a ultralow temperature (0.6 K) and low fields in their EuZn$_2$As$_2$ crystal ($n \sim 4.6 \times 10^{20}$ cm$^{-3}$), which led them to claim the presence of field-induced spin reorientation based on their MR analysis [30]. Their recent quantum oscillation studies have further revealed quantum-limited phenomena in EuZn$_2$As$_2$, suggesting nontrivial topology in its hole bands [60].

Similar to EuCd$_2$As$_2$ and EuCd$_2$P$_2$, AFM-EuZn$_2$As$_2$ also exhibits a significant NLAHE. In Wang *et al*'s work, the NLAHE contributes to 83% of the total Hall resistivity at $T_N$ [29]. Yi et al conducted a detailed investigation into the NLAHE in EuZn$_2$As$_2$, attributing the nonzero Berry curvature to the existence of FM short-range correlations [55]. Their results from electron spin resonance measurements reveal that this short-range FM order persists below $T_N$. The interplay between the long-range AFM order and the short-range FM order might explain the additional shoulder feature observed in the NLAHE curve below 12 K [55].

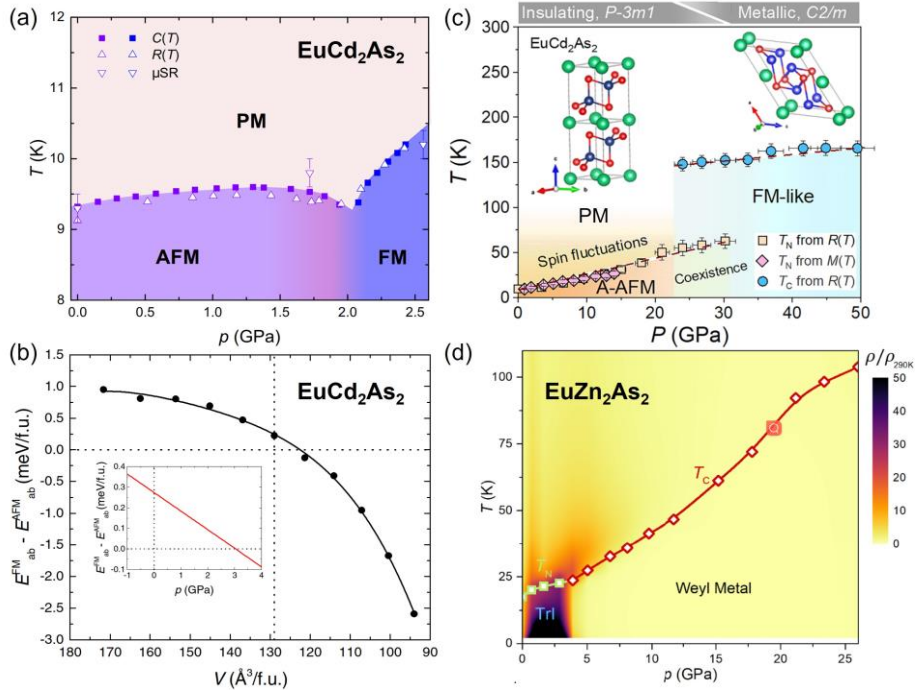

**Figure 9.** (a) Temperature-pressure ($T$-$p$) phase diagram of EuCd$_2$As$_2$. Reprinted from [61]. (b) Equation of state energy difference vs volume ($V$) between the fully relaxed FM and A-type AFM states ($E_{FM,ab}$−$E_{AFM,ab}$), where in both cases the moment point is in-plane to the nearest-neighbor Eu. The inset shows an enlarged view around the AFM$_{ab}$-to-FM$_{ab}$ transition region with the $x$ axis converted to pressure ($p$). Reprinted from [61]. (c) $T$-$p$ phase diagram for intrinsically insulating AFM-EuCd$_2$As$_2$. Reproduced from [50]. (d) $T$-$p$ phase diagram of EuZn$_2$As$_2$ at zero field, where the values of the normalized resistivity are color-coded. The empty rectangles and diamonds respectively represent $T_N$ and $T_C$ (determined from the local maxima in $d\rho/dT$). Reprinted from [31].

*5.2.4 Resistivity of AFM-EuZn$_2$P$_2$.* The resistivity data for AFM-EuZn$_2$P$_2$, as reported by Singh et al and Krebber et al, are depicted in figures 8(a) and 8(b), respectively [33,52]. Initially, EuZn$_2$P$_2$ was characterized as an insulator exhibiting A-type AFM ordering at 23.5 K. According to the Arrhenius model, the compound displays an energy band gap of approximately 0.2 to 0.3 eV under the zero-field condition [32,52]. In line with theoretical predictions and paralleling the behavior of other AFM-Eu$M_2X_2$ compounds [28], the gap narrows when the spins are realigned by an external field, as illustrated in the inset of figure 8(a). Despite this narrowing, the insulating behavior in EuZn$_2$P$_2$ persists even under fields as high as 14 T. Recently, Krebber et al succeeded in synthesizing semiconducting EuZn$_2$P$_2$ using a graphite crucible, and the resistivity curves are displayed in figure 8(b) [33]. At 100 K, the resistivity of their crystals is roughly six orders of magnitude lower than previously reported values, indicating unintentional heavy doping. Above 200 K, a metallic-like temperature dependence emerges, accompanied by a CMR of −10$^3$% at $T_N$ under a 4 T field.

*5.3 Pressure effect*

Pressure is a powerful tool for modulating electronic states. Several theoretical and experimental investigations have been carried out on Eu$M_2X_2$ materials, with a predominant focus on EuCd$_2$As$_2$ [31,50,61-67]. Gati et al and Du et al conducted hydrostatic pressure (up to 2.5 GPa) studies utilizing EuCd$_2$As$_2$ crystals with a bad metal behavior, indicative of a relatively high carrier concentration. Their findings were similar [61,66]. Primarily, the ground state of EuCd$_2$As$_2$ transitions from A-type AFM order, featuring in-plane spins (AFM$_{ab}$), to FM order, also with in-plane spins (FM$_{ab}$), at a critical pressure of 2 GPa. The phase diagram from Gati et al is shown in figure 9(a). This AFM-to-FM transition under pressure underscores the existence of FM interactions in AFM-EuCd$_2$As$_2$, consistent with experimental observations reported elsewhere. Furthermore, the $T_C$ rises with increasing pressure, suggesting an enhancement of FM coupling. In addition, the low-temperature resistivity exhibits a rapid increase under pressure when EuCd$_2$As$_2$ is in its AFM state, reverting to metallic behavior upon transformation into the FM state. Hence, the resistivity forms an insulating dome under intermediate pressures ranging from approximately 1.0 to 2.0 GPa, accompanied by a CMR effect up to –10$^5$% upon application of a magnetic field in this pressure region [66]. Simultaneously, $T_N$ undergoes a subtle variation, initially rising before declining. The changes in resistivity and $T_N$ imply alterations to the electronic structures. Lastly, Gati *et al*'s calculations under pressure suggest FM$_{ab}$ state becomes favored with decreasing cell volume, as shown in figure 9(b). Upon further pressure escalation, the FM$_{ab}$ state evolves into a FM state with spins aligned along the $c$ axis (FM$_c$), with their calculations proposing a critical pressure of 23 GPa for this transformation [61].

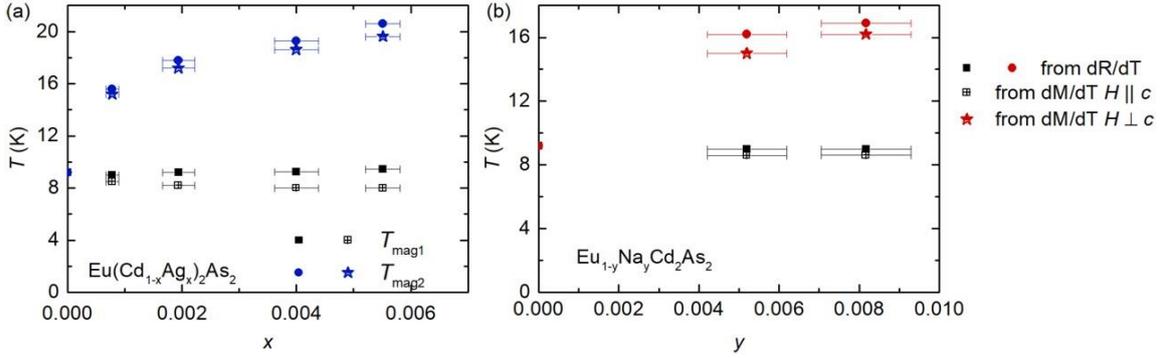

**Figure 10.** Temperature-doping ($T$-$x$) phase diagrams for Ag-substituted EuCd$_2$As$_2$ (a) and Na-substituted EuCd$_2$As$_2$. The two magnetic transitions, $T_{mag1}$ and $T_{mag2}$, are obtained from $dM/dT$ and $dR/dT$. Reprinted from [68].

Yu *et al* also conducted a pressure study, extending the pressure range up to 30 GPa [63]. Their findings corroborate the AFM-to-FM transition under pressure, enriching the phase diagram established by Gati *et al* and Du *et al* [61,66]. Notably, $T_C$ of EuCd$_2$As$_2$ reaches about 30 K at a pressure of 10 GPa. Moreover, Yu *et al* observed a butterfly-shaped MR curve at pressures exceeding 25 GPa, attributed to the magnetic hysteresis of EuCd$_2$As$_2$ in the FM state [63]. And they claimed a FM$_c$ state at fields above 19 GPa. It is noteworthy that the EuCd$_2$As$_2$ samples used by Yu *et al* appear to be more semiconducting compared to those in the experiments by Gati *et al*'s and Du *et al*'s experiments [61,63,66]. Subsequently, Jose *et al* contributed an additional phase diagram, encompassing pressures up to 42.8 GPa [65]. At this extreme pressure, the $T_C$ value determined via time-domain synchrotron Mössbauer spectroscopy exceeds 80 K and the increase in $T_C$ shows no sign of saturation. Contrary to Gati *et al*'s prediction [61], Jose *et al* discovered that the magnetic moments predominantly orient within the *ab* plane, even at pressures as high as 42.8 GPa [65].

Chen *et al* carried out another high-pressure study on EuCd$_2$As$_2$, pushing the pressure envelope to 50 GPa [50], as illustrated in figure 9(c). A distinguishing feature of their work is the use of an intrinsically insulating sample for experimentation. Intriguingly, they discovered that the AFM$_{ab}$ state endures up to 30 GPa, beyond which a FM-like state emerges at pressures above 24 GPa. Remarkably, this FM-like state exhibits a $T_C$ reaching 150 K at 24.0 GPa, which increases linearly with pressure at a rate of 0.69 K/GPa. Additionally, Chen *et al* observed a novel structural transition in EuCd$_2$As$_2$ from the $P$-3$m$1 to the $C$2/$m$ space group above 28.2 GPa [50], which is unprecedented in previous studies. The discrepancies between Chen *et al*'s findings and prior research might suggest that carrier density plays an important role in the establishment of FM order. This is reminiscent of the FM-Eu$M_2X_2$ compounds synthesized via the salt flux method [36,38,39], which will be elaborated on in Section 6.

The magnetic and transport properties of EuZn$_2$As$_2$ under high pressure have been investigated by Luo *et al* [31]. Analogous to EuCd$_2$As$_2$, the application of moderate pressure renders EuZn$_2$As$_2$ increasingly insulative, leading to a huge CMR. Upon reaching a pressure of 4 GPa, EuZn$_2$As$_2$ undergoes a transition from an AFM semiconductor to a FM metal. The temperature-pressure phase diagram elucidated by Luo *et al* is depicted in figure 9(d). $T_C$ for EuZn$_2$As$_2$ in the FM state reaches 100 K under a pressure of 26.1 GPa, which is anticipated to increase further under higher pressures. Moreover, Rybicki *et al*'s investigation on EuZn$_2$P$_2$ shows that the energy gap is markedly suppressed by pressure, resulting in an insulator-to-metal transition [67]. They also documented a substantial rise in $T_N$, surpassing 40 K at a field of 9.5 GPa. It is noteworthy that the $T_N$ values reported by Rybicki *et al* at elevated pressures might actually represent $T_C$ values [67], as observed in other high-pressure studies.

The high-pressure investigations reveal that the pressure-induced FM order is a pervasive characteristic across the Eu$M_2X_2$ compounds. The magnetic properties under pressure appear to be intimately linked to the intrinsic carrier density within the crystal lattice. On the basis of the aforementioned studies, it appears that EuZn$_2$As$_2$ and EuZn$_2$P$_2$ exhibit a higher $T_C$ than EuCd$_2$As$_2$ at comparable pressures. This observation can be attributed to the reduced cell volumes of the Zn analogs, aligning with the $T_C$ trends observed in FM-Eu$M_2X_2$ compounds synthesized through the salt flux method [38]. To further elucidate these phenomena, future research calls for additional controlled experiments under high pressure on Eu$M_2X_2$ materials. These experiments should explore the effects of varying carrier densities and initial ground states to address outstanding questions and refine our understanding of the underlying mechanisms.

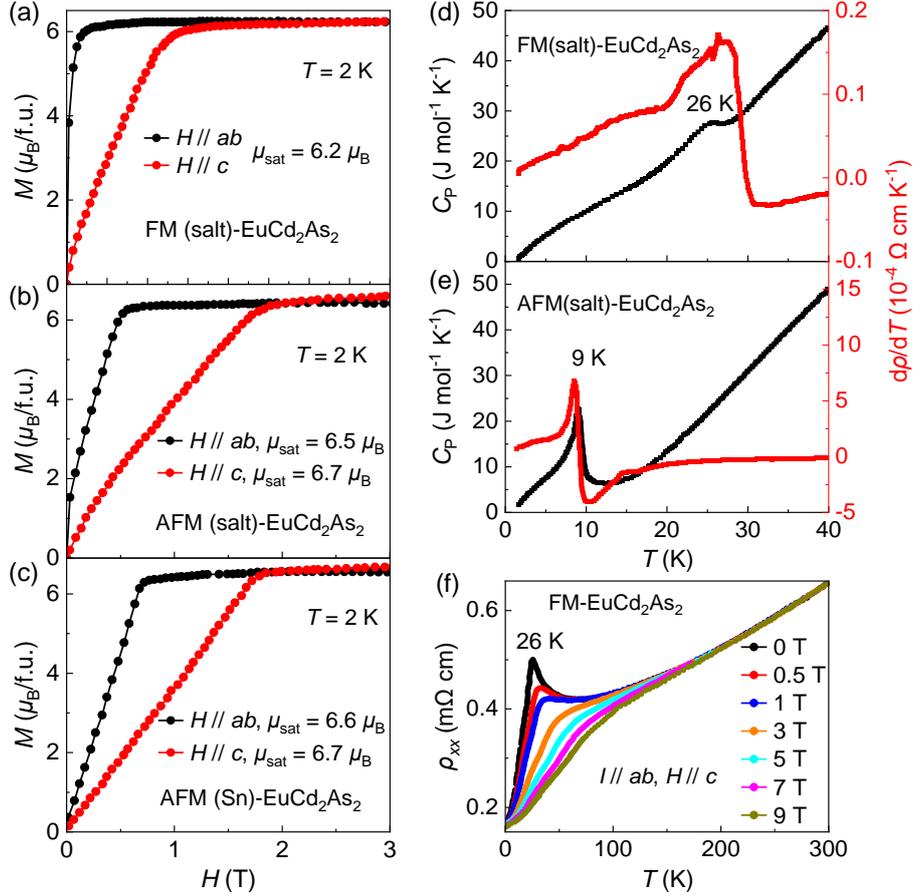

**Figure 11.** (a-c) Field-dependent magnetization with in-plane ($H // ab$) and out-of-plane ($H // c$) fields at 2 K for FM-EuCd$_2$As$_2$, salt flux grown AFM-EuCd$_2$As$_2$, and Sn flux grown AFM-EuCd$_2$As$_2$, respectively. (d-e) Temperature-dependent specific heat $C_P$ (black line, left axis) and resistivity derivatives d$\rho$/d$T$ (red line, right axis) for FM-EuCd$_2$As$_2$ and AFM-EuCd$_2$As$_2$, respectively. (f) Temperature-dependent of resistivity for FM-EuCd$_2$As$_2$ under various fields. Panels (a-e) are reproduced from [36]. Panel (f) is reproduced from [37].

### 5.4 Doping effect in AFM-EuCd$_2$As$_2$

To our knowledge, only two instances of intentional heterovalent doping in EuCd$_2$As$_2$ have been reported thus far: Kuthanazhi *et al* utilized Na$^+$ and Ag$^+$ substitutions to introduce hole carriers [68], whereas Nelson *et al* employed La$^{3+}$ substitution to generate extra electrons [69]. Despite endeavors to increase the carrier concentration by enhancing the initial ratio of dopant elements, the achieved doping levels remain fairly low. For instance, a nominal 20% Na content in the starting mixture (Eu$_{0.8}$Na$_{0.2}$Cd$_2$As$_2$) yields a mere 0.8% effective doping (Eu$_{0.992}$Na$_{0.008}$Cd$_2$As$_2$), a situation similarly encountered with Ag substitution at Cd sites and La substitution at Eu sites. These difficulties in doping may stem from the hexagonal close-packed structure of Eu$M_2X_2$ compounds coupled with their high bonding energies.

Kuthanazhi *et al* observed a splitting of the original AFM transition in EuCd$_2$As$_2$ into two distinct transitions [68]. The lower transition, labeled as $T_{mag1}$, is independent of doping, whereas the higher transition, denoted as $T_{mag2}$, increase gradually with the increased chemical substitution. This behavior was noted for both Ag- and Na-doped samples, as illustrated in figure 10. The authors found that $T_{mag2}$ is associated with a FM component of moments within the *ab* plane, leading them to posit that a shift in band filling could potentially stabilize a FM phase in EuCd$_2$As$_2$. This hypothesis is corroborated by their theoretical calculations and aligns with the FM ground state observed in defect-rich Eu$M_2X_2$ compounds. It is noteworthy that, in Kuthanazhi *et al*'s study, the fluctuations in $T_{mag2}$ and the Weiss temperatures ($\theta_W$), indicators of FM interaction strength, are modest, even as the chemical doping level varies by an order of magnitude [68]. This observation may suggest that carrier concentration is not the predominant determinant of FM correlation strength, which will be revisited in Section 6.2.2.

Nelson *et al* prepared *n*-type EuCd$_2$As$_2$ crystals through La doping at Eu sites, achieving a dopant concentration of 30 ppm and a doping level about $10^{17}$ cm$^{-3}$ [69]. Their ARPES measurements, performed at 6 K on potassium-dosed samples, unveiled a clear band gap for AFM-EuCd$_2$As$_2$. This finding provides strong evidence supporting the recent reports of semiconducting behavior in EuCd$_2$As$_2$ by Santos-Cottin *et al*, as previously introduced [25].

Although heterovalent doping is challenging in Eu$M_2X_2$, it holds promise as a method for controlling carrier concentration and tuning the physical properties of these materials, warranting further exploration.

# 6. Physical properties of FM-Eu*M*$_2$*X*$_2$

Recent investigations have demonstrated that all four compounds in the Eu$M_2X_2$ series ($M$ = Zn, Cd; $X$ = P, As), which exhibit a FM ground state, can be synthesized using the salt flux method. This section aims to delineate the magnetic and transport properties of these FM-Eu$M_2X_2$ materials and to discuss the mechanisms underlying the interlayer FM couplings.

## 6.1 FM-EuCd$_2$As$_2$

*6.1.1 Magnetism and resistivity.* Initial interest in EuCd$_2$As$_2$ in a FM state was sparked by theoretical predictions that its out-of-plane ferromagnetism, i.e., the *c*-axis polarized spin state, can host an idea WSM phase with a single pair of Weyl nodes [11,13,14]. However, while FM correlations are detected above and below $T_N$ (9.5 K) in AFM-EuCd$_2$As$_2$ [10,53], no net FM component is discernible when the long-range AFM order is established. Surprisingly, Jo *et al* managed to synthesize FM-EuCd$_2$As$_2$ with an ordering temperature of 26 K merely by altering the flux from Sn to a mixture of NaCl and KCl [36]. This represents a rare case of effectively manipulating the ground state of a material through such a simple modification.

Figure 11(a)-(c) summarize the magnetization data for AFM- and FM-EuCd$_2$As$_2$. Jo *et al* found that both AFM and FM variants of EuCd$_2$As$_2$ can be synthesized using the salt flux method, contingent upon the concentration of Eu in the solution. Interestingly, the magnetic properties, encompassing the $T_N$ values and the saturation fields, of AFM-EuCd$_2$As$_2$ crystals grown via the salt flux technique align closely with those grown using the Sn flux method. All EuCd$_2$As$_2$ samples exhibit an easy-plane magnetic anisotropy, signifying that the Eu spins lie in the *ab* plane. It is worth noting that the saturation field of FM-EuCd$_2$As$_2$ for *H* // *ab* is noticeably reduced, indicating a pronounced enhancement of the FM coupling between the Eu layers. This analogous decrease in saturation field under in-plane fields is also observed in other FM-Eu$M_2X_2$ compounds, which will be elaborated in Section 6.2.1. The specific heat and resistivity derivative collectively substantiate the bulk FM transition in FM-EuCd$_2$As$_2$, rather than signaling a mere FM component arising from a canted AFM order. Additional corroborative evidence, such as the direct visualization of FM domains through magneto-optical imaging, further attests to the presence of FM order in FM-EuCd$_2$As$_2$. High-energy X-ray diffraction experiments, along with benchtop X-ray techniques, have revealed a slight deficiency in Eu content (approximately 1% to 4%). This Eu deficiency is believed to play a critical role in a band shift of the hole pocket, thereby affecting the material's ground state properties.

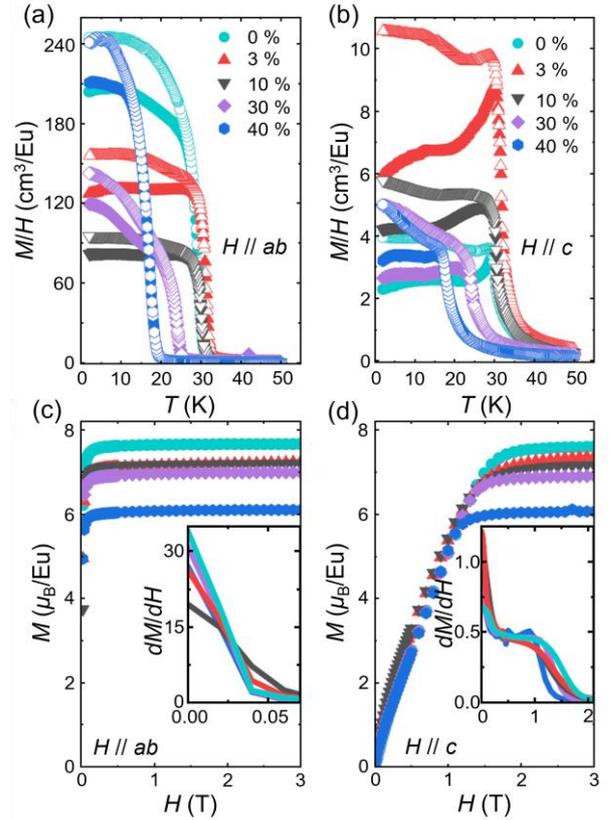

**Figure 12.** Temperature-dependent magnetization of Eu$_{1-x}$Ba$_x$Cd$_2$As$_2$ with a field of 50 Oe applied (a) in the *ab* plane and (b) along the *c* axis. Closed symbols represent ZFC data, and open symbols represent FC data. Magnetization field sweeps for field applied (c) in-plane and (d) along the *c* axis. Insets in (c) and (d) show the field derivative (*dM/dH*) for the two field orientations. Reprinted from [70].

Roychowdhury *et al* studied the charge transport of FM-EuCd$_2$As$_2$ under various magnetic fields, as shown in figure 11(f) [37]. The material exhibits metallic conductivity, with the exception of an upturn observed above the $T_C$ of FM ordering. Moreover, the transition peak at 26 K is rapidly

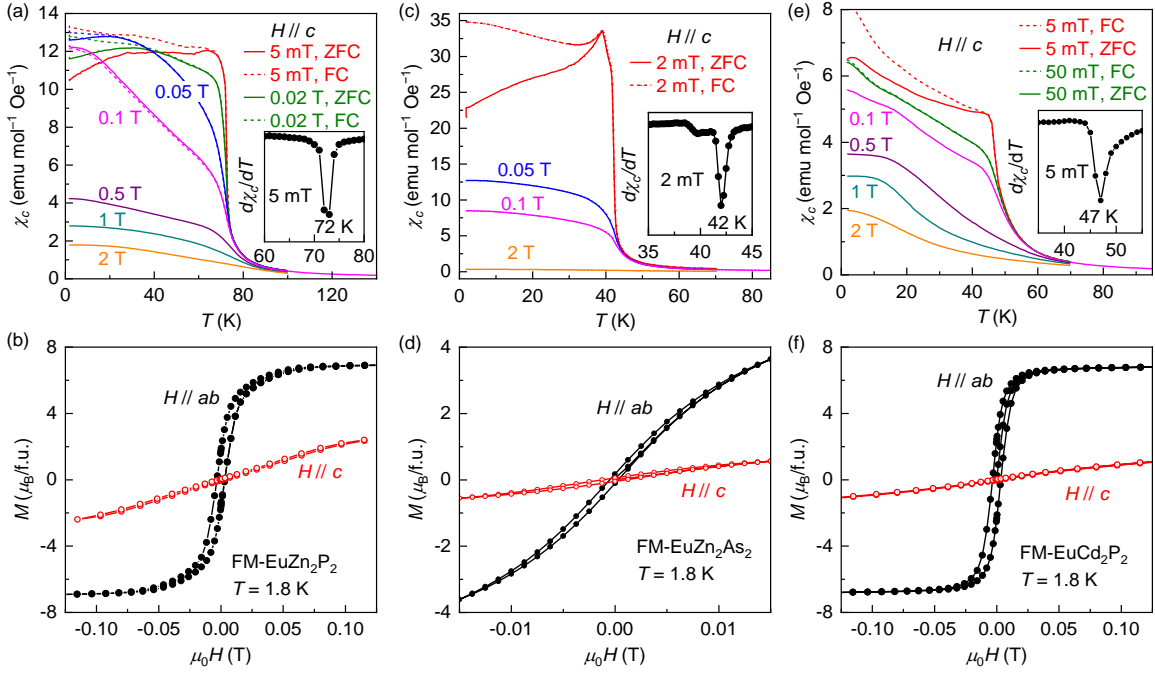

**Figure 13.** Temperature-dependent susceptibility and magnetic hysteresis loops of FM-EuZn$_2$P$_2$ [(a), (b)], FM-EuZn$_2$As$_2$ [(c), (d)], and FM-EuCd$_2$P$_2$ [(e), (f)] under various conditions. The insets in panels (a), (c), and (e) depict the transition temperatures as determined by the derivative of the susceptibility data. Reproduced from [38,39].

suppressed by applied fields, resulting in a pronounced nMR effect.

*6.1.2 Doping effect.* Theoretical calculations by Wang *et al* predicted that replacing half of the Eu atoms with Ba in AFM-EuCd$_2$As$_2$ would induce a lattice expansion, thereby stabilizing a FM ground state with an out-of-plane spin configuration [13]. To test this hypothesis, Sanjeewa *et al* synthesized a series of Eu$_{1-x}$Ba$_x$Cd$_2$As$_2$ single crystals using the salt flux method, with Ba doping concentrations ranging from 0 to 40% [70]. The magnetization data for these crystals are presented in figure 12. The splitting of zero-field-cooled (ZFC) and field-cooled (FC) data, alongside the markedly enhanced in-plane susceptibility, unambiguously signals the in-plane FM order in the Ba-doped samples, in agreement with the findings reported by Jo *et al* [36]. Several key observations merit attention:

Firstly, with low Ba substitution levels, $T_C$ remains almost invariant; however, a significant decline in $T_C$ is observed at higher doping concentrations, indicative of a weakening in the FM coupling strength.

Secondly, contrary to expectations of monotonic evolution, the susceptibility magnitude initially decreases before reverting to the value of the parent compound, suggestive of non-monotonic change in the spin canting angle.

Thirdly, the authors did not detect any Eu vacancies in their undoped FM-EuCd$_2$As$_2$, contradicting earlier reports [36]. Yet, a subtle carrier concentration shift can provoke substantial alterations in the magnetism of Eu$M_2X_2$ compounds, potentially exceeding the detection limits of the experimental apparatus.

Fourthly, the theoretical prediction that Ba doping induces ferromagnetism in EuCd$_2$As$_2$ could not be confirmed in this study, as the parent compound was already FM. To properly assess this prediction, semiconducting EuCd$_2$As$_2$ should be serve as the parent material.

*6.2 FM-EuCd$_2$P$_2$, FM-EuZn$_2$P$_2$, and FM-EuZn$_2$As$_2$*

*6.2.1 Magnetism.* The aforementioned results clearly demonstrate that FM-EuCd$_2$As$_2$ was successfully synthesized using the salt flux method. The alteration in its magnetic ground state is attributed to the reduced Eu$^{2+}$ content and the consequent shift in the Fermi level. Given that other members of the Eu$M_2X_2$ family share the CaAl$_2$Si$_2$-type structure, exhibit an A-type AFM configuration, and display similar short-range FM correlations, it is logical to explore the potential for discovering additional FM compounds within this family. Indeed, our recent investigations have revealed that FM single crystals of EuZn$_2$P$_2$ ($T_C$ = 72 K), EuZn$_2$As$_2$ ($T_C$ = 42 K), and EuCd$_2$P$_2$ ($T_C$ = 47 K) can also be grown utilizing the salt flux technique [38,39].

Figure 13 shows the susceptibility ($H$ // $c$) and magnetization ($H$ // $ab$ and $H$ // $c$) of the FM variants of EuZn$_2$P$_2$, EuZn$_2$As$_2$, and EuCd$_2$P$_2$. Clear bifurcations in the FC and ZFC data, the significant increase around $T_C$, and the presence of magnetic hysteresis loops unequivocally confirm the FM ground state in these three materials. This contrasts with the sharp peak characteristic of AFM phases at $T_N$ and their non-hysteretic magnetization behaviors. The $T_C$ values were determined through the derivative of the susceptibility, $d\chi_c/dT$, yielding results of 72 K for FM-EuZn$_2$P$_2$, 42 K for FM-EuZn$_2$As$_2$, and 47 K for FM-EuCd$_2$P$_2$. The in-plane magnetization curves for all three compounds display a pronounced hysteresis effect, whereas the out-of-plane hysteresis loops are notably weak, suggesting that the magnetic easy axis lies within the $ab$ plane. Furthermore, the in-plane saturation fields for the FM-Eu$M_2X_2$ series are markedly lower when compared to those depicted in the bottom panels of figure 3, indicating a substantially enhanced FM coupling between the Eu planes. Additionally, the Weiss temperatures ($\theta_W$) obtained from fitting (78.8 K for FM-EuZn$_2$P$_2$, 49.2 K for FM-EuZn$_2$As$_2$, 50.2K for FM-EuCd$_2$P$_2$) are closely aligned with the respective $T_C$ values, which further supports the conclusion that EuZn$_2$P$_2$, EuZn$_2$As$_2$, and EuCd$_2$P$_2$ are intrinsic ferromagnets rather than canted antiferromagnets.

Base on the single-crystal X-ray diffraction data, the crystal structures of FM-Eu$M_2X_2$ are virtually identical to those of their AFM counterparts, except for a slight proportion of Eu vacancies in the FM phases (~5% for FM-EuZn$_2$P$_2$, ~0.2% for FM-EuCd$_2$P$_2$) [38,39], reminiscent of the situation observed in EuCd$_2$As$_2$. We have previously discussed that AFM interactions in Eu$M_2X_2$ family is mediated by the superexchange mechanism via the Eu-$X$-$X$-Eu path [52]. Given that the Eu-$X$ and $X$-$X$ bond lengths and Eu-$X$-$X$ bond angles remain essentially constant, it is reasonable to infer that the interlayer AFM interaction persists. Therefore, the emergent ferromagnetism must stem from an enhancement of interlayer FM coupling, which now dominates, a condition that can only be attributed to the presence of low-concentration hole carriers. The induction of ferromagnetism by a low carrier concentration is not exclusive to the Eu$M_2X_2$ series. Precedents abound, including FM semiconductors Eu$X$ ($X$ = O, S, Se, Te) [71,72], the diluted magnetic semiconductor (Ga,Mn)As [73,74], the Mn pyrochlore Tl$_{2-x}$Sc$_x$Mn$_2$O$_7$ [75], and other low-carrier-density ferromagnets such as Ca$_{1-x}$La$_x$B$_6$, EuB$_6$, UTeS, PbSnMnTe, etc [76-79].

*6.2.2 Characteristic temperatures of Eu$M_2X_2$.* FM-EuZn$_2$P$_2$ attains a remarkably high $T_C$ of 72 K among Eu-based materials, nearly three times that of FM-EuCd$_2$As$_2$, which has a $T_C$ of 26 K. This raises the natural question of what factors critically determine the magnitude of $T_C$. The characteristic temperatures ($T_C$, $T_N$, $\theta_W$) for both FM- and AFM-Eu$M_2X_2$ are plotted in figure 14, as a function of the Eu-layer distances ($d_{inter}$), i.e., the $c$-axis values listed in Table 1. Surprisingly, it is discovered that both the FM transition temperature $T_C$ and the Weiss temperature $\theta_W$ exhibit a linear correlation with the interlayer distance. This linear dependence strongly suggests that the FM ordering temperatures of the FM-Eu$M_2X_2$ series are heavily reliant on the strength of interlayer Eu-Eu coupling. It is not unexpected that such a linear relationship does not hold in the AFM phases, given the distinct mechanisms governing AFM interactions. Moreover, we observe that the $T_C$ values are consistently somewhat lower than their corresponding $\theta_W$. This disparity can likely be attributed to the existing influence of AFM interactions. Consequently, it is unsurprising to find a greater divergence between $T_C$ and $\theta_W$ for FM-EuZn$_2$P$_2$ and FM-EuZn$_2$As$_2$, owing to their shorter Eu-$X$-$X$-Eu path.

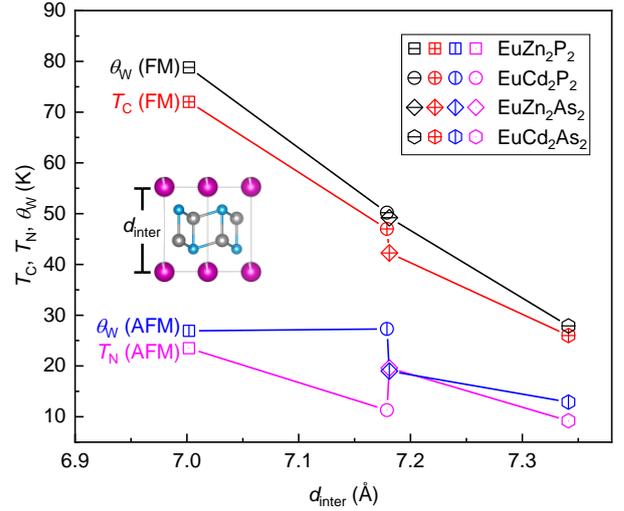

**Figure 14.** The characteristic temperatures ($T_C$, $T_N$, $\theta_W$) of Eu$M_2X_2$ ($M$ = Zn, Cd; $X$ = P, As) as a function of the Eu-layer distance, i.e., the $c$ axis. The squares, circles, diamonds, and hexagons represent the data of FM EuZn$_2$P$_2$, EuCd$_2$P$_2$, EuZn$_2$As$_2$, and EuCd$_2$As$_2$, respectively. Reprinted from [38].

Another critical aspect that requires discussion is the role of carrier densities. It has been established that the hole carriers, induced by Eu vacancies, play a crucial role in the establishment of FM order. However, unlike the scenario observed in electron-doped FM semiconductors Eu$X$ ($X$ = O, S, Se) [80-82], the final $T_C$ appears not to be significantly impacted by variations in carrier density. This inference is backed by several observations. Firstly, $T_C$ values reported for FM-EuCd$_2$As$_2$ across various research groups exhibit good consistency [36,37,70]. Secondly, our attempts to grow Eu$M_2X_2$ crystals under differing flux concentrations resulted in virtually unaltered $T_C$ values. Thirdly, as previously discussed, the $T_C$ of FM-EuCd$_2$As$_2$ derived from heterovalent doping does not exhibit a pronounced dependency on the doping concentration [68]. These insights suggest the existence of a threshold concentration of Eu vacancies necessary for the carrier-induced transition from AFM to FM order. For instance, the carrier concentration of FM-EuCd$_2$P$_2$ (~4.6 × 10$^{19}$ cm$^{-3}$, equivalent to 0.26% Eu vacancy/f.u.) is about an order of magnitude higher than that of AFM-EuCd$_2$P$_2$ (3.6 × 10$^{18}$ cm$^{-3}$, equivalent to 0.02% Eu

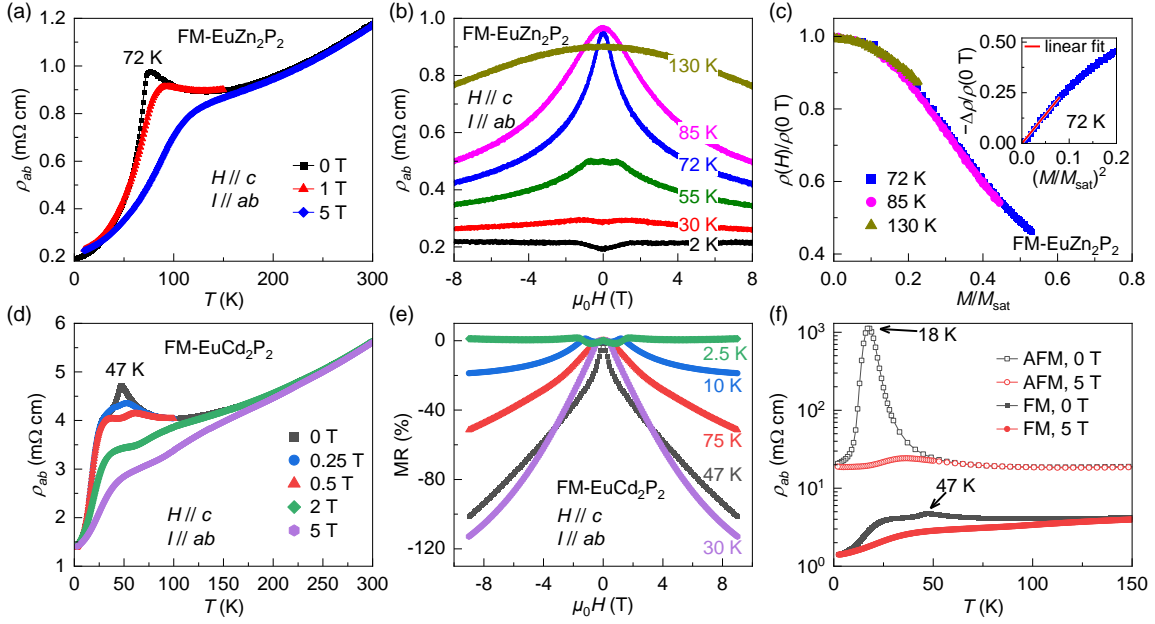

**Figure 15.** Electrical transport data of FM-EuZn$_2$P$_2$ (a-c, reproduced from [38]) and FM-EuCd$_2$P$_2$ (d-f, reproduced from [39]). (a) In-plane resistivity of FM-EuZn$_2$P$_2$ under several fields along the $c$ axis. (b) Field-dependent resistivity at various temperatures from 2 to 130 K. (c) Normalized resistivity $\rho(H)/\rho(0\ T)$ at 72 K (blue squares), 85 K (magenta circles), and 130 K (dark yellow triangles) plotted against normalized field-induced magnetization $M/M_{sat}$, where $M_{sat}$ is the saturated magnetization. (d) In-plane resistivity of FM-EuCd$_2$P$_2$ under several fields along the $c$ axis. (e) Calculated MR of FM-EuCd$_2$P$_2$ as a function of field at several representative temperatures. (f) A comparison of the resistivity behaviors between AFM- and FM-EuCd$_2$P$_2$. Note that the resistivity is plotted on a logarithmic scale.

vacancy/f.u.) [38,39]. The threshold concentration falls somewhere between these two values. Once the hole concentration surpasses this threshold, Eu$M_2X_2$ transitions into the FM state with a characteristic $T_C$, without manifesting any intermediate magnetic phase. This threshold warrants further investigation, given the current inability to continuously modulate the Eu vacancy concentration in samples using the Sn or salt flux growth methods. Future theoretical and experimental efforts should be devoted to elucidating the precise impact of carrier density on $T_C$ and $T_N$.

*6.2.3 Charge transport.* The holes induced by Eu vacancies not only alter the magnetic order of Eu$M_2X_2$ compounds, but also dramatically decrease their resistivity. The charge transport properties of FM-EuZn$_2$P$_2$ and FM-EuCd$_2$P$_2$ are illustrated in figure 15, whereas data for FM-EuZn$_2$As$_2$ are omitted due to the inferior crystal quality, which hindered the data collection. In stark contrast to the AFM counterparts, both FM-EuZn$_2$P$_2$ and FM-EuCd$_2$P$_2$ display metallic behavior, accompanied by a resistivity peak at $T_C$. By applying the magnetic field, magnetic scatterings are significantly diminished, resulting in the rapid suppression of the resistivity peak and the manifestation of a pronounced nMR effect around $T_C$. Panel (b) depicts the field-dependent resistivity, of FM-EuZn$_2$P$_2$ at various temperatures, while the calculated MR as a function of field for FM-EuCd$_2$P$_2$ is displayed in panel (e). A slight rise in $\rho(H)$ and MR($H$) was observed under low fields when the temperatures are well below $T_C$, attributed to the enhanced canting of spins toward the $c$ axis in the field. Panel (c) plots the normalized resistivity of FM-EuZn$_2$P$_2$, $\rho(H)/\rho(0\ T)$, at 72 K ($T_C$), 85 K (1.2 $T_C$), and 130 K (1.8 $T_C$) versus normalized magnetization ($M/M_{sat}$). The consistent trend across the three datasets suggests that the MR of FM-EuZn$_2$P$_2$ is closely tied to its magnetization, signifying that magnetic scattering is a predominant factor in the MR effect. By plotting $-\Delta\rho/\rho(0\ T)$ as a function of $(M/M_{sat})^2$, we discerned that their relationship could be described by the scaling function $-\Delta\rho/\rho(0\ T) = C(M/M_{sat})^2$ [83]. The derived coefficient $C$ is consistent with the prediction of the Majumdar-Littlewood model [38]. Hence, the nMR of FM-EuZn$_2$P$_2$ can potentially be augmented merely by reducing the carrier concentration.

In addition, FM-EuCd$_2$P$_2$ exhibits a hump in resistivity around 30 K, which is notably absent in other FM-Eu$M_2X_2$ materials. This anomaly may stem from the competition between short-range magnetic correlations and interlayer FM coupling below $T_C$, given the pronounced magnetic fluctuations in AFM-EuCd$_2$P$_2$ and the subtle modifications to the crystal structure upon transitioning between the two phases. Figure 15(f) presents the contrasting resistivity magnitudes of the AFM- and FM-EuCd$_2$P$_2$. The striking disparity in the resistivity behavior and magnetic properties between the AFM and FM states, coupled with the subtle alteration in carrier density, positions EuCd$_2$P$_2$ as a promising candidate for future spintronic applications.

## 7. Conclusion and perspectives

This work offers a comprehensive review of Eu$M_2X_2$ ($M$ = Zn, Cd; $X$ = P, As) materials, delving into their crystal structure, magnetic properties, electrical transport characteristics, and electronic structure in detail. The initial impetus for studying Eu$M_2X_2$ stemmed from the theoretical prediction of a WSM state in EuCd$_2$As$_2$, although this concept has faced significant scrutiny. Nevertheless, meticulous exploration of Eu$M_2X_2$ properties has unveiled a rich landscape ripe with exotic phenomena such as the NLAHE, CMR effect, and highly modifiable magnetic and electrical transport behaviors. From a practical standpoint, the tunable properties of Eu$M_2X_2$ materials make them exceptionally promising candidates for spintronic applications. Our review concludes with several key takeaways:

(a) *Eu$M_2X_2$ ($M$ = Zn, Cd; $X$ = P, As) are most likely magnetic semiconductors lacking a topologically nontrivial nature.* Previous claims designating EuCd$_2$As$_2$ primarily relied on theoretical calculations and experiments involving heavily *p*-doped EuCd$_2$As$_2$ samples. These claims have been systematically challenged by recent meticulous theoretical studies and experiments conducted on ultraclean or *n*-doped EuCd$_2$As$_2$ crystals [25-28]. Indeed, experimental phenomena once attributed to Weyl physics, such as the giant NLAHE, could be accounted for by alternative mechanisms [16,55].

(b) *Despite hosting an A-type AFM order, FM correlations predominate within the Eu$M_2X_2$ family.* This peculiar confluence engenders a constellation of distinctive properties across Eu$M_2X_2$ compounds, including strong FM fluctuations, the CMR effect, the NLAHE, carrier-induced FM ordering, and pressure-induced FM ordering. Furthermore, the intensity of FM interactions is intimately tied to the dimensions of the unit cell, as evidenced by the $T_C$ trends observed in Eu$M_2X_2$ synthesized via the salt flux method [36,38], and the enhancement of $T_C$ under elevated pressures [31,50,65].

(c) *The narrow band gap and its sensitivity to the spin configurations jointly contribute to the highly tunable charge transport properties in Eu$M_2X_2$.* Reported modulation techniques encompass the application of magnetic field, pressure, bias currents [27], carrier doping, and strain [62,64], etc. The high tunability lays the groundwork for the prospective utilization of Eu$M_2X_2$ in various applications.

(d) *A phase diagram is compiled to depict the correlation between carrier concentration and the resultant magnetism and charge transport characteristics in Eu$M_2X_2$ compounds.* As shown in figure 16, three distinct regions emerge: A yellow-shaded area signifies AFM semiconducting phases of Eu$M_2X_2$, characterized by an exceedingly low hole density (typically below $10^{17}$ cm$^{-3}$) and a pronounced shift in resistivity magnitude. A blue-shade area signifies Eu$M_2X_2$ exhibiting relatively low resistivity for intermediate carrier densities (ranging approximately from $10^{17}$ cm$^{-3}$ to $10^{20}$ cm$^{-3}$), with a subtle variation in resistivity magnitude, yet maintaining A-type AFM magnetic order. A black-shaded area represents FM-Eu$M_2X_2$ manifesting typical metallic behavior for higher carrier concentrations within the crystal lattice (approximately exceeding $10^{20}$ cm$^{-3}$). Note that for the heavily *p*-doped sample with a carrier density of $10^{21}$ cm$^{-3}$, the estimated Eu vacancy, based on a single-band model, amounts to merely 0.06 per formula unit. Besides, we notice that the carrier densities of EuCd$_2$As$_2$, synthesized via different methods, span a broad range of five orders of magnitude, from $10^{15}$ to $10^{20}$ cm$^{-3}$, which accounts for the diverse properties reported in literature for EuCd$_2$As$_2$. This phase diagram implies that a subtle fluctuation in the concentration of Eu defects could precipitate a dramatic alteration in the magnetic and charge transport properties of Eu$M_2X_2$ compounds.

In the past few years, considerable research efforts have been dedicated to exploring Eu$M_2X_2$ compounds, yielding significant advancements. However, several outstanding questions still need addressing, and further manipulation of these materials' properties is warranted. Below, we would like to outline some potential research avenues in this specific topic:

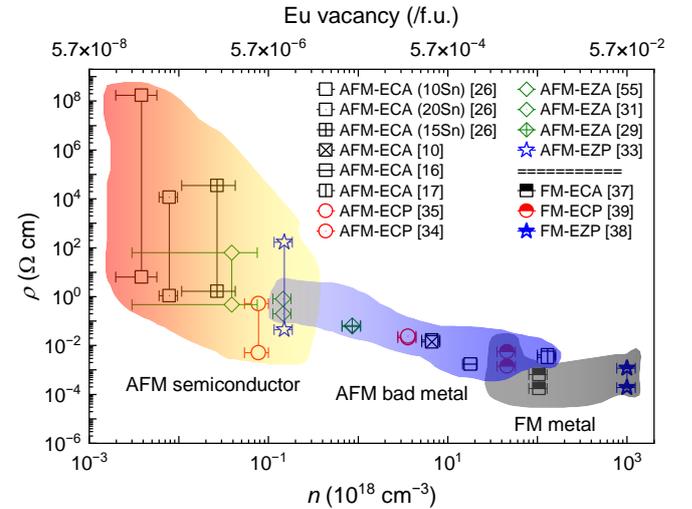

**Figure 16.** The upper bound and lower bounds of resistivity for Eu$M_2X_2$ compounds, plotted as a function of the carrier densities (bottom axis). Corresponding value of Eu vacancy is indicated on the top axis. Each distinct symbol represents a compound in either the AFM or FM ground state. Resistivity for each phase is denoted by dual data points, which represent the maximum and minimum values extracted from the resistivity curves without the field. Carrier densities were calculated using the OHE contributions from the Hall resistivity curves. The horizontal error bars are estimated based on the variability of carrier density across different

temperatures. Where Hall data at multiple temperatures are unavailable, an error bar equivalent to 25% of the carrier density is assumed. ECA, ECP, EZA, and EZP designate the compounds EuCd$_2$As$_2$, EuCd$_2$P$_2$, EuZn$_2$As$_2$, and EuZn$_2$P$_2$, respectively. The sources of the data are denoted within the figure.

(e) *The mechanism by which long-range FM order is established in EuM$_2$X$_2$ compounds.* Experimental findings have demonstrated that a modest application of pressure or a slight adjustment in carrier density can induce a switch in the magnetic ground state [38,61]. However, a comprehensive theoretical framework to explain this mechanism is currently absent. The specific pathway through which interlayer FM coupling is mediated is yet to be elucidated.

(f) *The origins of the giant NLAHE and CMR.* These phenomena are commonly observed in EuM$_2$X$_2$ compounds, suggesting a universe physical cause. Although several mechanisms have been postulated for specific materials [16,42,55], a comprehensive understanding that encapsulates all members of EuM$_2$X$_2$ family has yet to be fully elucidated.

(g) *The continuous modulation of carrier density in EuM$_2$X$_2$ compounds.* In Section 6.2.2, we posit the existence of a critical carrier density threshold for the transition between AFM and FM state. However, this hypothesis remains unverified, as the salt flux method can not finely tune the carrier density within EuM$_2$X$_2$ crystals. Moreover, chemical doping presents significant challenges in this system. Electrostatic gating emerges as a promising solution, suitable not only for EuM$_2$X$_2$ thin films [84-86] but also for micrometer-scale single-crystalline lamellae [87]. It would be intriguing to determine whether the $T_N$ and $T_C$ vary continuously with changes in carrier density, or if a sudden jump occurs instead. And it is also of great interest to see the maximum $T_C$ attainable through the modulation of carrier density, which is a critical consideration for the application of EuM$_2$X$_2$ in spintronics.

(h) *Enhance $T_C$ of FM-EuM$_2$X$_2$ through the reduction of interlayer Eu-Eu distances.* As shown in figure 14 in Section 6.2.2, a linear increase in $T_C$ is observed as the interlayer spacing decreases. Moreover, the highest $T_C$ values obtained for EuCd$_2$As$_2$ through high-pressure studies exceed 150 K at 49.5 GPa [50], whereas the record for EuZn$_2$As$_2$ stands at 100 K at 26.1 GPa [31]. These findings suggest that $T_C$ could be further enhanced by compressing the unit cell volume via chemical pressure, specifically through doping with ions of smaller size. Substituting Eu sites with small divalent elements, such as Ca$^{2+}$ and Yb$^{2+}$, could effectively strengthen interlayer FM coupling. However, non-magnetic dopants may conversely diminish magnetic interactions, as evidenced by the $T_C$ trend of Eu$_{1-x}$Ba$_x$Cd$_2$As$_2$ [70]. Gd$^{3+}$ is also a viable dopant candidate due to its smaller radius and the same $4f^7$ electron configuration. Moreover, Gd$^{3+}$ allows for the introduction of extra electrons into the lattice to explore the properties of *n*-type EuM$_2$X$_2$.

(i) *High-pressure studies involving both insulating and metallic variants of EuM$_2$X$_2$.* Previous high-pressure experiments conducted on both insulating and metallic EuCd$_2$As$_2$ have unveiled notable discrepancies [50,65]. This discrepancy has yet to be satisfactorily explained. It holds considerable intrigue to probe whether other EuM$_2$X$_2$ compounds, characterized by differing carrier densities, might exhibit a spectrum of pressure-induced phenomena. Furthermore, among the quartet, EuZn$_2$P$_2$ may potentially display the highest $T_C$ under pressure, attributed to its notably shorter interlayer Eu-Eu distance.

(j) *Enhance the CMR effect.* The CMR effect is a pivotal phenomenon due to its critical importance in magnetic memory and sensing technologies. Based on observations of the magnetoresistive responses in EuM$_2$X$_2$ compounds, it appears that more insulating behavior correlates with a greater CMR effect. Therefore, enhancing the CMR effect could be achieved through a moderate increase in the energy gap.

(k) *Explore the possibility of manipulating the ground state in other Eu-based materials with a CaAl$_2$Si$_2$-type structure.* As mentioned in Section 2, numerous Eu-based compounds exhibiting an A-type AFM ordering were synthesized previously, such as EuAl$_2$Ge$_2$ [44], EuMg$_2$(Sb/Bi)$_2$ [45,88], Eu(Zn/Cd)$_2$Sb$_2$ [89,90], and EuMn$_2$X$_2$ (X = P, As, and Sb) [4,46,91], etc. It would be intriguing to investigate whether FM order could be induced in these materials.

(l) *The peculiar features of EuCd$_2$P$_2$.* Among the EuM$_2$X$_2$ compounds, EuCd$_2$P$_2$ stands out as particularly distinctive. It exihibits exceptionally strong FM fluctuations. Notably, its resistivity peak is significantly elevated above $T_N$, with the peak temperature itself being correlated to the carrier density within the sample. Additionally, FM-EuCd$_2$P$_2$ displays a pronounced resistivity hump around 30 K, as shown in figure 15(d). These anomalous characteristics remain unexplained and warrant further investigative efforts.

In summary, even without a topologically nontrivial nature, the EuM$_2$X$_2$ family (M = Zn, Cd; X = P, As) remains a rich field for exploring the intricate interplay between magnetism and transport properties.

## Acknowledgements

This work was supported by the National Natural Science Foundation of China (Grants No. 12204094), the Natural Science Foundation of Jiangsu Province (Grant No. BK20220796), the Start-up Research Fund of Southeast University (Grant No. RF1028623289), the Interdisciplinary program of Wuhan National High Magnetic Field Center (WHMFC) at Huazhong University of Science and Technology (Grant No. WHMFC202205).